\documentclass[aps, prd,twocolumn, amsmath, floats,floatfix,
superscriptaddress, nofootinbib,noshowpacs]{revtex4-1}

\usepackage{stackengine}
\usepackage{array}
\usepackage{ifpdf}
\ifpdf \usepackage[pdftex]{graphicx} \graphicspath{{figures/pdf/}}
\DeclareGraphicsExtensions{.pdf} \else \usepackage[dvips]{graphicx}
\usepackage{graphicx} \usepackage{pstricks} \usepackage{pst-node}
\usepackage{pst-blur} \graphicspath{{figures/eps/}}
\DeclareGraphicsExtensions{.eps} \definecolor{Pink}{rgb}{1.,0.75,0.8}
\definecolor{cyan2}{cmyk}{0.40,0,0,0} \fi
\usepackage{tikz}
\usetikzlibrary{positioning}
\usetikzlibrary{arrows}

\usepackage{float}

\usepackage{csquotes}

\usepackage{tabularx}
\usepackage{array}
\newcolumntype{Z}{>{\centering\let\newline\\\arraybackslash\hspace{0pt}}X}

\usepackage{bbm}
\usepackage{xspace}

\usepackage{accents}

\usepackage{siunitx}

\usepackage[normalem]{ulem}

\usepackage[colorlinks=true]{hyperref}
\hypersetup{
  citecolor  = blue
}

\def\diag{\operatorname{diag}}

  \def\eps{\epsilon}

\def\ET{{\tt Einstein Toolkit}\xspace}
\def\SENR{{\tt SENR/NRPy+}\xspace}
\def\NRPy{{\tt NRPy+}\xspace}
\def\RNS{{\tt Hydro\_RNSID}\xspace}
\def\GRHydro{{\tt GRHydro}\xspace}

\def\p0{\partial_0}

\def\apj{Astrophys. J.}  \def\aap{Astron. Astrophys.}
\def\mnras{Mon. Not. R. Astron. Soc.}  \def\prd{Phys. Rev. D}
\def\prl{Phys. Rev. Lett.}  
\def\apjl{Astrophys. J. Letters}
\def\apjs{Astrophys. J. Supplement Series}

\usepackage{tikz}
\usepackage{amsmath} \usepackage{amssymb} \usepackage{listings}
\usepackage{mathtools} \usepackage{subfigure}
\usepackage{empheq}

\newcommand\exactfbox[1]{\fbox{\hskip1em#1\hskip1em}}

\begin{document}

\title{Numerical relativity in spherical coordinates: A new dynamical spacetime
and general relativistic MHD evolution framework for the Einstein Toolkit}

\author{
Vassilios Mewes}
\email{mewesv@ornl.gov}
\affiliation{National Center for Computational Sciences, Oak Ridge National Laboratory, P.O. Box 2008, Oak Ridge, TN 37831-6164, USA}
\affiliation{Physics Division, Oak Ridge National Laboratory, P.O. Box 2008, Oak Ridge, TN 37831-6354, USA}
\affiliation{Center for Computational Relativity and Gravitation, and School of Mathematical Sciences, Rochester Institute of Technology, 85 Lomb Memorial Drive, Rochester, New York 14623, USA}
\author{
Yosef Zlochower
}
\affiliation{Center for Computational Relativity and Gravitation, and School of Mathematical Sciences, Rochester Institute of Technology, 85 Lomb Memorial Drive, Rochester, New York 14623, USA}
\author{
Manuela Campanelli
}
\affiliation{Center for Computational Relativity and Gravitation, and School of Mathematical Sciences, Rochester Institute of Technology, 85 Lomb Memorial Drive, Rochester, New York 14623, USA}
\author{
Thomas W. Baumgarte
}
\affiliation{Department of Physics and Astronomy, Bowdoin College, Brunswick, Maine 04011, USA}
\author{
Zachariah B. Etienne
}
\affiliation{Department of Physics and Astronomy, West Virginia University, Morgantown, West Virginia 26506, USA}
\affiliation{Center for Gravitational Waves and Cosmology, West Virginia University, Chestnut Ridge Research Building, Morgantown, West Virginia 26505, USA}
\author{
Federico G. Lopez Armengol
}
\affiliation{Center for Computational Relativity and Gravitation, and School of Mathematical Sciences, Rochester Institute of Technology, 85 Lomb Memorial Drive, Rochester, New York 14623, USA}
\affiliation{Instituto Argentino de Radioastronom\'ia (IAR), C.C. No. 5, 1894 Buenos Aires, Argentina}
\author{
Federico Cipolletta
}
\affiliation{INFN-TIFPA, Trento Institute for Fundamental Physics and Applications, Via Sommarive 14, I-38123 Trento, Italy}
\affiliation{Dipartimento di Fisica, Universit\`a di Trento, Via Sommarive 14, I-38123 Trento, Italy}
\affiliation{Center for Computational Relativity and Gravitation, and School of Mathematical Sciences, Rochester Institute of Technology, 85 Lomb Memorial Drive, Rochester, New York 14623, USA}

\begin{abstract}
We present {\tt SphericalNR}, a new framework for the publicly available
{\tt Einstein Toolkit} that numerically solves the
Einstein field equations coupled to the equations of general relativistic
magnetohydrodynamics (GRMHD) in a 3+1 split of spacetime in spherical coordinates
without symmetry assumptions. The spacetime evolution is performed using
reference-metric versions of either the Baumgarte-Shapiro-Shibata-Nakamura equations
or the fully covariant and conformal Z4 system with constraint damping.
We have developed a reference-metric version of the {\it Valencia formulation} of GRMHD
with a vector potential method, guaranteeing the absence of magnetic
monopoles during the evolution. In our framework, every dynamical field (both spacetime and matter)
is evolved using its components in an orthonormal basis with respect to the spherical
reference-metric. Furthermore, all geometric information
about the spherical coordinate system is encoded in source terms appearing in
the evolution equations. This allows for the straightforward extension of
Cartesian high-resolution shock-capturing finite volume codes to use
spherical coordinates with our framework. To this end, we have adapted
\GRHydro, a Cartesian finite volume GRMHD code already available in the
{\tt Einstein Toolkit}, to use spherical coordinates. We present the full
evolution equations of the framework, as well as details of its implementation
in the {\tt Einstein Toolkit}. We validate {\tt SphericalNR} by demonstrating it passes a
variety of challenging code tests in static and dynamical spacetimes.
\end{abstract}

\pacs{ 04.25.D-, 
  04.30.-w, 
  04.70.Bw, 
  95.30.Sf, 
  97.60.Lf 
}

\maketitle

\section{Introduction}
\label{sec:introduction}

The detection of gravitational waves (GW) from binary black hole (BBH)
mergers via the ground-based LIGO and VIRGO detectors~\citep{Abbott:2016blz,
TheLIGOScientific:2016wfe,Abbott:2016nmj,Abbott:2017vtc,Abbott:2017gyy,Abbott:2017oio}
and the simultaneous detection of GW and electromagnetic (EM) radiation from
binary neutron star (BNS) mergers~\cite{TheLIGOScientific:2017qsa,
GBM:2017lvd,Monitor:2017mdv} has opened a new window into the Universe.
Accurate numerical simulations of compact binary mergers are crucial for
estimating the physical parameters of detected systems~\cite{Aasi2014}, and
for informing physical models about the evolution of matter at nuclear densities
in the postmerger remnant of BNSs and BHNSs~\cite{MetzgerKilonovaLRR2017,RosswogMMA2015}.
However, the full self-consistent numerical simulation of a compact object merger through
coalescence and subsequent postmerger evolution at high resolution
is an extremely challenging computational problem involving vast differences in
length and timescales, as well as different approximate symmetries during
the stages of the evolution. In particular, during the inspiral and merger, the absence of
approximate axisymmetry lends itself to the use of Cartesian coordinates,
while the postmerger remnant has approximate symmetries that are better
captured by using spherical coordinates.

In the field of numerical relativity it is now possible to do
self-consistent simulations of compact object binary mergers:
The first general relativistic hydrodynamics (GRHD) BNS merger simulation~\cite{ShibataBNS2000},
the breakthrough simulations of BBH mergers~\cite{Pretorius2005,Campanelli2006,Baker2006},
the first GRHD black hole neutron star (BHNS) merger simulation~\cite{Shibata2006},
the first general relativistic magnetohydrodynamics (GRMHD) BNS merger
simulations~\cite{HAD2008,Liu2008}, and the first GRMHD simulation of BHNS
mergers~\cite{ChawlaBHNS2010}.
Self-consistent simulations of these multimessenger sources requires
the accurate modeling of the dynamical spacetime evolution and
magnetohydrodynamic (MHD) flows within and near compact objects.
To this end, many codes have been written that solve the equations
of GRMHD. Generally such codes fall
into two categories: GRMHD codes coupled to a 3+1 dynamical spacetime
solver (see e.g.~\cite{SACRA2005,IllinoisGRMHD2005,ValenciaGRMHD2006,HAD2006,
Whisky2007,CerdaDuran:2008pv,X-ECHO2011,GRHydro2014,IGM2015,WhiskyGRRMHD2013,SpECTRE2017,
Spec_mag2018,FambriAder2018,Cipolletta:2019geh}), and GRMHD
codes that adopt analytical expressions for the spacetime geometry -- which could be either
exact, if the spacetime is stationary, or approximate for some dynamical
spacetimes (see e.g.~\cite{Koide1999,Villiers2003,HARM2003,Cosmos++2005,Komissarov2004,
NobleHARM2006,BonazzolaGRMHD2007,NobleHarm2009,Athena++2016,BHAC2017,
H-AMR2019}).
In general, the codes coupled to a 3+1 spacetime solver perform the fluid
evolution on Cartesian or multipatch grids, while many of the fixed background spacetime
codes employ curvilinear coordinates. A notable exception are the codes
of~\cite{CerdaDuran:2008pv} and~\cite{X-ECHO2011}, which solve the fluid equations in
curvilinear coordinates, coupled to a dynamical spacetime solver in an approximate,
constrained evolution formulation of the Einstein
field equations~\cite{CerdaDuranCFC+2005,CorderoCarrion2009,CorderoCarrion2012}, which
is a generalization of the conformal flatness condition~\cite{Wilson1996,Isenberg2008}.

Often, numerical error in conservation of momentum is smallest in the direction of coordinate
lines.  Accordingly, codes written in Cartesian coordinates conserve linear momentum well, while
codes using spherical coordinates conserve angular momentum well.
In GRMHD, momenta are only conserved when spacetime (rather than just coordinate)
symmetries are present, due to the appearance of source terms in the evolution
equations. Many astrophysical systems of interest to
multimessenger astrophysics possess a natural axisymmetry at first approximation,
so that one expects a better conservation
of angular momentum in spherical coordinates, which allow the azimuthal
coordinate to be aligned with the direction of this symmetry.
Examples include (see references within the cited review articles):
core-collapse supernovae~\cite{JankaCoreCollapse2007,BurrowsCorecollapse2013},
compact binary merger remnants~\cite{ShibataBHNSLRR2011,FaberBNSLRR2012,
BaiottiBNS2017,DuezReview2019,Radice:2020ddv},
pulsars~\cite{PulsarsLRR2005}, magnetars~\cite{TurollaMagnetars2015,Mereghetti:2015asa,KaspiMagnetars2017},
and self-gravitating accretion disks~\cite{AbramowiczAccretionDisksLRR2013}.
While there are techniques to mitigate the nonconservation
of angular momentum in Cartesian coordinates (see
e.g.~\cite{CallAdvection2010,Mignone2012,Byerly2014}), it would
generally be more desirable to use numerical grids with
spherical sampling, representing all tensors and vectors in the spherical basis.
Evolving Einstein's field equations in spherical coordinates introduces
both conceptual and computational problems associated with coordinate singularities,
but many of these issues have been resolved in recent years.

Among the formalisms of Einstein's field equations most commonly used in numerical
simulations is the Baumgarte-Shapiro-Shibata-Nakamura (BSSN) formulation \cite{Shibata1995,Baumgarte1999}.  Choices
made in the original version of this formulation are suitable in Cartesian coordinates
only, but a generalization involving a reference-metric formalism allows for applications
in any coordinate system (see, e.g.,
\cite{Bonazzola2004,Shibata2004,GourgoulhonEric2007,BrownCovariant2009,Montero2012b}).
In the absence of spherical symmetry, coordinate singularities can be handled by
properly rescaling components of tensors \cite{Baumgarte2013,Montero2014,Baumgarte2015},
which is equivalent to expressing all tensor components in a frame that is orthogonal with
respect to the reference-metric.  In~\cite{fCCZ42014}, the formalism was extended to the
Z4 formalism (see, e.g., \cite{BonaZ42003,BernuzziZ4c2010,Alic2012}).  The \SENR code
\cite{Ruchlin2018} provides a flexible
computational framework for the implementation of the formalism in a broad class of
coordinate systems.  Using this framework, we previously implemented this approach
in the \ET \cite{Mewes:2018szi}.

In this work we extend the framework presented in~\cite{Mewes:2018szi}
to evolve the GRMHD equations in a reference-metric formalism, and add
a constraint-damping formulation for the spacetime evolution to the existing
framework. Our new code applies this strategy by coupling the GRMHD equations with a fully dynamical
spacetime evolution via the BSSN or fully covariant and conformal Z4 (fZZC4)
system~\cite{fCCZ42014}
on three-dimensional spherical coordinate grids (i.e., no symmetry assumptions are made).
The GRMHD evolution equations are evolved using a vector potential method.
We also use the moving-puncture gauge conditions~\cite{Campanelli2006,Baker2006}.
To our knowledge, {\tt SphericalNR} is the first framework solving the
coupled BSSN/fCCZ4 and GRMHD equations in three-dimensional spherical coordinate
grids without symmetry assumptions.

The paper is organized as follows: In Sec.~\ref{sec:GRMHD}, we describe the
evolution equations for both spacetime and GRMHD in spherical coordinates.
In Sec.~\ref{sec:SphericalGRHydro} we describe
the implementation of the GRMHD reference-metric
evolution formalism in the \ET~\cite{ETK_2019_3522086}. Code tests are presented in
Sec.~\ref{sec:code_tests}. Finally, Sec.~\ref{sec:conclusions} contains conclusions and
discussions.
Throughout this paper and in the code implementation we use
geometrized and rationalized (geometrized Heaviside-Lorentz) units in
which $c=G=M_{\odot}=\epsilon_0=\mu_0=1$, where $c$, $G$, $M_{\odot}$,
$\epsilon_0$, and $\mu_0$ are the speed of light, gravitational constant, solar mass,
vacuum permittivity and vacuum permeability,
respectively. Latin indices denote spatial indices, running from 1 to 3;
Greek indices denote spacetime indices, running from 0 to
3 (0 is the time coordinate); and the Einstein summation convention is used.

%
\section{Basic Equations: Dynamical spacetime and GRMHD equations in spherical coordinates}
\label{sec:GRMHD}
%

%
\subsection{Spacetime evolution in spherical coordinates}
\label{subsec:fCCZ4}
%

We dynamically evolve the gravitational fields using a numerical relativity framework in
spherical coordinates implemented using the \ET infrastructure (see \cite{Mewes:2018szi}).
Our framework builds upon a reference-metric formulation \cite{BrownCovariant2009,GourgoulhonEric2007,Montero2012b}
of the BSSN formalism~\cite{Nakamura1987,Shibata1995,Baumgarte1999}.
We appropriately scale out singular factors from components of tensors so that, for nonsingular
spacetimes, all numerically evolved variables remain regular even at the origin and on the polar axis
(see \cite{Baumgarte2013,Baumgarte2015}). In this paper we extend the spacetime evolution
code described in~\cite{Mewes:2018szi} to include the
fCCZ4 formalism (see \cite{fCCZ42014}), which applies the same reference-metric
formalism and rescaling approach to the CCZ4 evolution
equations~\cite{Alic2012,Alic2013}. This represents a conformal
reformulation of the constraint-damped Z4
system (see \cite{BonaZ42003,BonaZ42004,GundlachZ42005}; see also \cite{BernuzziZ4c2010} for
an alternative conformal reformulation of Z4). We have now implemented the fCCZ4 formalism
in the {\tt SphericalNR} framework, and therefore provide, as a reference, key equations below.

The constraint-damped Z4 system~\cite{BonaZ42003,BonaZ42004,GundlachZ42005} replaces
Einstein's equations by
\begin{eqnarray} \label{Z4}
R_{\mu \nu} &+& \nabla_{\mu} \mathcal{Z}_{\nu} + \nabla_{\nu} \mathcal{Z}_{\mu}
- \frac{\kappa_1}{\alpha} [ n_{\mu} \mathcal{Z}_{\nu} + n_{\nu} \mathcal{Z}_{\mu} \nonumber \\
&-& (1+\kappa_2)g_{\mu \nu} n_{\lambda} \mathcal{Z}^{\lambda} ]
= 8 \pi (T_{\mu \nu} - \frac{1}{2} g_{\mu \nu} T),
\end{eqnarray}
where $R_{\mu \nu}$ is the (spacetime) Ricci tensor, $g_{\mu \nu}$ the spacetime metric,
$\nabla_\mu$ its associated covariant derivative,
$T_{\mu \nu}$ the stress-energy tensor, $T \equiv g^{\mu \nu} T_{\mu \nu}$ its
trace, $\mathcal{Z}_{\nu}$ a four-vector of constraints, and $\alpha$ the lapse function.
We will shortly associate
the timelike vector $n^\mu$ with the normal on spatial slices.
Finally, $\kappa_1$ (units of inverse length) and
$\kappa_2$ (dimensionless) are two damping coefficients,
and all nonconstant constraint related modes are damped when
$\kappa_1 > 0$ and $\kappa_2 > -1$~\cite{GundlachZ42005}.

Following the discussion in~\cite{Alic2013} regarding the stability of the
evolution system in the presence of black holes (BH), we have
redefined $\kappa_1 \to \kappa_1/\alpha$.
The Z4 system reduces to the
Einstein equations when the constraint vector $Z_{\mu}$ vanishes.

We start with a $3+1$ split of spacetime (see \cite{Darmois1927}) and foliate
the four-dimensional spacetime with a set of nonintersecting
spacelike hypersurfaces $\Sigma$.  We denote the future-pointing, timelike normal
on $\Sigma$ as $n_\mu$, and refer to the projection of the spacetime metric
$g_{\mu\nu}$ onto $\Sigma$ as the spatial metric
\begin{equation} \label{spatial_metric}
\gamma_{\mu \nu} = g_{\mu \nu} + n_\mu n_\nu.
\end{equation}
Expressing the normal vector in terms of a lapse function $\alpha$ and a shift vector $\beta^i$,
\begin{equation} \label{eq::normal}
n^{\mu} = \left(\frac{1}{\alpha},-\frac{\beta^i}{\alpha} \right)
\end{equation}
or
\begin{equation} \label{eq::normalcov}
n_{\mu} = (-\alpha,0),
\end{equation}
we may write the four-dimensional line element as
\begin{eqnarray}
ds^2 &=& g_{\mu \nu}dx^{\mu}dx^{\nu} \nonumber \\
&=& -\alpha^2 dt^2 + \gamma_{ij}(dx^i+\beta^i dt)(dx^j+\beta^j dt),
\end{eqnarray}
and the spacetime metric $g_{\mu \nu}$ as
\begin{equation}\label{3p1_metric}
 g_{\mu \nu} = \left( \begin{array}{cc}
  -\alpha^2 + \beta_i \beta^i & \beta_j \\
  \beta_i & \gamma_{i j}
 \end{array}\right).
\end{equation}

As in the BSSN formalism we conformally rescale the spatial metric according to
\begin{equation}
\bar{\gamma}_{ij} = e^{-4\phi} \gamma_{ij},
\end{equation}
where $\bar \gamma_{ij}$ is the conformally related metric and $e^\phi$ the conformal
factor.   The latter can be written as
\begin{equation}
e^{4\phi} = (\gamma / \bar{\gamma})^{1/3},
\end{equation}
where $\gamma$ and $\bar{\gamma}$ are the determinants of the physical and conformally
related metric, respectively.  We will assume that
\begin{equation}
\partial_t \bar{\gamma} = 0,
\label{eq:lagrangian_choice}
\end{equation}
meaning that $\bar{\gamma}$ remains equal to its initial value throughout the evolution.  This
choice, referred to as the \enquote{Lagrangian} choice in \cite{BrownCovariant2009}, simplifies
some expressions in particular in the context of the GRMHD evolution, as explained below.
We also rescale the trace-free part of the extrinsic curvature according to
\begin{equation}
\bar{A}_{ij} = e^{-4\phi}\left(K_{ij}-\frac{1}{3}\gamma_{ij}K\right),
\end{equation}
where $K_{ij}$ is the physical extrinsic curvature and $K \equiv \gamma^{ij}K_{ij}$ its trace.

The central idea of the reference-metric formalism\footnote{Splitting the metric into
background and departures from the
background (which need not to be small) is also done in
bimetric formalisms~\cite{Rosen1963,Cornish1964,Rosen1973,NahmadAchar1987,Katz1985,
Katz1988,Katz1990} in general relativity, in which reference-metrics are
employed to give physical meaning to pseudotensors in curvilinear coordinates.  We emphasize
that we do not consider extensions of general relativity here; rather, we use the reference-metric
only as a convenient approach to express Einstein's equations
(see also \cite{Gourgoulhon1994,Bonazzola2004,Shibata2004,Cook2008}).}
is to express the conformally
related metric as the sum of a background metric $\hat \gamma_{ij}$ and deviations $h_{ij}$
(which need not to be small),\footnote{Strictly speaking, it is sufficient to introduce a reference connection
only (e.g.~\cite{BrownCovariant2009}), but it is convenient to assume that this connection is associated
with a reference-metric.}
\begin{equation}
\bar{\gamma}_{ij} = \hat{\gamma}_{ij} + h_{ij}.
\end{equation}
For our purposes it is particularly convenient to choose as the reference-metric the flat metric
in spherical coordinates,
\begin{equation}
\label{backMetDef}
\hat{\gamma}_{ij} = \begin{pmatrix}
1 & 0 & 0 \\
0 & r^2 & 0 \\
0 & 0 & r^2 {\rm sin}^2 \theta
\end{pmatrix}.
\end{equation}
Another key ingredient is evolving vector and tensor components in the
orthonormal basis with respect to $\hat{\gamma}_{ij}$ instead
of components in the spherical coordinate basis. To this end, we introduce a
set of basis vectors $\hat{\mathbf{e}}^{\{k\}}_{i}$
that are orthonormal with respect to the background metric $\hat \gamma_{ij}$:
\begin{equation}\label{eq:hatted_basis_vectors}
  \hat{\gamma}_{ij} = \delta_{\{k\}\{l\}} \hat{\mathbf{e}}^{\{k\}}_{i} \hat{\mathbf{e}}^{\{l\}}_{j}.
\end{equation}
Since $\hat{\gamma}_{ij}$ is diagonal, the orthonormal vector basis tetrad
and its inverse are given by
\begin{align}
  \hat{\mathbf{e}}^{\{k\}}_{i} & = \diag (1,r, r \sin \theta), \\
  \hat{\mathbf{e}}^{k}_{\{i\}} & = \diag (1,1/r, 1/(r \sin \theta)),
\end{align}
where we have adopted a notation involving plain Latin and
Latin indices surrounded with curly braces: The components of a
tensor ${\bf T}$ in the standard coordinate basis will be denoted
using the former, while the tensor components in the background
orthonormal basis will be denoted by the latter, respectively.
In this notation, we may write the deviation tensor $h_{ij}$ as
\begin{equation}
h_{ij} = \hat{\mathbf{e}}^{\{k\}}_{i} \hat{\mathbf{e}}^{\{l\}}_{j} h_{\{k\}\{l\}} 
\end{equation}
and similarly write $\bar{A}_{ij}$ as\footnote{This is a novel notation since, in standard
references, different symbols are used for rescaled quantities. For instance,
in~\cite{Baumgarte2013}, the pairs $\{ h_{ij}, h_{\{i\}\{j\}}\}$,
$\{\bar{A}_{ij}, \bar{A}_{\{k\}\{l\}}\}$ are denoted as
$\{\epsilon_{ij},h_{ij}\}$, $\{\bar{A}_{ij}, a_{ij}\}$, respectively.}
\begin{equation} \label{aij}
\bar{A}_{ij} = \hat{\mathbf{e}}^{\{k\}}_{i} \hat{\mathbf{e}}^{\{l\}}_{j} \bar{A}_{\{k\}\{l\}}.
\end{equation}
While we will write most equations in terms of coordinate components (i.e.~indices without curly
braces), the code uses components in the orthogonal basis (i.e.~with curly indices) as dynamical
variables.

As in the original BSSN formalism, we introduce conformal
connection functions $\bar{\Lambda}^i$ as independent
variables.  In the context of the reference-metric formalism the
$\bar \Lambda^i$ satisfy the initial constraint
\begin{equation}
\bar{\Lambda}^i - \Delta \Gamma^i = 0,
\end{equation}
where
\begin{equation}
\Delta \Gamma^i \equiv \bar{\gamma}^{jk}\Delta\Gamma^i_{jk}
\end{equation}
and
\begin{equation}
\Delta \Gamma^i_{jk} \equiv \bar{\Gamma}^i_{jk} - \hat{\Gamma}^i_{jk}.
\end{equation}
Contrary to the Christoffel symbols themselves, differences between Christoffel
symbols transform as rank-3 tensors.  We compute the
$\Delta \Gamma^i_{jk}$ from
\begin{align} \label{Delta_Gamma}
\Delta \Gamma^i_{jk} &= \frac{1}{2} \bar{\gamma}^{il}
\left(\hat{\mathcal{D}}_j \bar{\gamma}_{kl}
+\hat{\mathcal{D}}_k \bar{\gamma}_{jl}
- \hat{\mathcal{D}}_l \bar{\gamma}_{jk} \right) \nonumber \\
&= \frac{1}{2} \bar{\gamma}^{il}
\left(\hat{\mathcal{D}}_j h_{kl}
+\hat{\mathcal{D}}_k h_{jl}
- \hat{\mathcal{D}}_l h_{jk} \right),
\end{align}
where $\hat{\mathcal{D}}_i$ is the covariant derivative associated with
the reference-metric $\hat \gamma_{ij}$,
and where we have used $\hat{\mathcal{D}}_i \hat{\gamma}_{jk} =0$ in the
second equality.
Derivatives of coordinate components of tensors are evaluated by
using the chain rule to analytically take derivatives of the
basis vectors $\hat{\mathbf{e}}^{\{k\}}_{i}$, while
the orthonormal components are finite-differenced numerically in the
code, e.g.
\begin{equation}
\partial_k h_{ij} = \hat{\mathbf{e}}^{\{l\}}_{i} \hat{\mathbf{e}}^{\{m\}}_{j} \partial_k h_{\{l\}\{m\}}
 + h_{\{l\}\{m\}} \partial_k \left(\hat{\mathbf{e}}^{\{l\}}_{i} \hat{\mathbf{e}}^{\{m\}}_{j} \right).
\end{equation}
Similar to our treatment of the metric and extrinsic curvature we
write
\begin{equation} \label{Lambda}
\bar{\Lambda}^i = \hat{\mathbf e}^i_{\{j\}} \bar{\Lambda}^{\{j\}}
=\begin{pmatrix}
\bar{\Lambda}^{\{r\}} \\
\bar{\Lambda}^{\{\theta\}} / r \\
\bar{\Lambda}^{\{\varphi\}} / (r \sin \theta)
\end{pmatrix}
\end{equation}
and evolve the orthonormal components $\bar{\Lambda}^{\{i\}}$ in our code.

One of the attractive features of the reference-metric formalism is that all
quantities, including the conformal connection functions $\bar{\Lambda}^i$,
transform as tensor densities of weight zero\footnote{
A tensor density of weight $w$ acquires a power $J^w$ in a coordinate transformation,
where $J\equiv \mathrm{det} |J^{i'}_j|$ is the determinant of the Jacobian
matrix of the coordinate transformation $J^{i'}_j\equiv \frac{\partial x^{i'}}{\partial x^j}$.
For example, a rank (2,0) tensor density of weight $w$ transforms as:
$\mathcal{T}^{i' j'} = J^w J^{i'}_k J^{j'}_l \mathcal{T}^{kl}$.
Tensor densities of weight zero therefore transform as ordinary (or absolute)
tensors with familiar coordinate transformations.}
(see~\cite{BrownCovariant2009}).

We now extend the above formalism to the $Z4$ formulation, following~\cite{fCCZ42014}.
We start with a 3+1 decomposition
of the constraint vector $\mathcal{Z}_\mu$,
\begin{equation}
\mathcal{Z}_\mu = g_{\mu}^{~\nu} \mathcal{Z}_\nu = \gamma_\mu^{~\nu} \mathcal{Z}_\nu - n_\mu n^\nu \mathcal{Z}_\nu,
\end{equation}
and define
\begin{eqnarray}
\Theta &\equiv& - n_{\lambda} \mathcal{Z}^{\lambda} = \alpha \mathcal{Z}^0,\\
Z_{i} &\equiv& \gamma_{i}^{\lambda} \mathcal{Z}_{\lambda}.
\end{eqnarray}
In Eq.~\eqref{LambdaReDef} below we will absorb the spatial parts $Z_i$ into the connection functions $\bar \Lambda^i$, but we will
evolve $\Theta$ as a new independent variable.

In order to write the evolution equations of the fCCZ4 system,
we first define a new tensor
\begin{equation}
\bar{R}^{Z4}_{ij} \equiv \bar{R}_{ij} + \mathcal{D}_i Z_j + \mathcal{D}_j Z_i,
\end{equation}
where $\mathcal{D}_i$ is the covariant derivative associated with the
spatial metric $\gamma_{ij}$ and $\bar{R}_{ij}$ is the Ricci tensor
associated with the conformal metric $\bar{\gamma}_{ij}$,
\begin{align}
\bar{R}_{ij} &=- \frac{1}{2} \bar{\gamma}^{kl} \hat{\mathcal{D}}_k \hat{\mathcal{D}}_l \bar{\gamma}_{ij}
+ \bar{\gamma}_{k(i}\hat{\mathcal{D}}_{j)}\Delta\Gamma^k + \Delta\Gamma^k \Delta\Gamma_{(ij)k}  \nonumber \\
&+ \bar{\gamma}^{kl} (2 \Delta\Gamma^m_{k(i}\Delta\Gamma_{j)ml}
+ \Delta\Gamma^m_{ik} \Delta\Gamma_{mjl}),
\end{align}
where parentheses around indices indicate the symmetric part of a
tensor: $T_{(ij)} \equiv \frac{1}{2}(T_{ij}+T_{ji})$.
We next define new conformal connection functions according to
\begin{equation} \label{LambdaReDef}
\tilde{\Lambda}^i \equiv \Delta\Gamma^i + 2 \bar{\gamma}^{ij} Z_j,
\end{equation}
where we have used a tilde in order to distinguish these objects from $\bar \Lambda^i$.
We then have
\begin{equation}
Z^i = \frac{1}{2} e^{-4\phi} (\tilde{\Lambda}^i - \Delta\Gamma^i).
\end{equation}
With these definitions, we may now write $R^{Z4}_{ij}$ as
\begin{align}
\bar{R}^{Z4}_{ij} = &- \frac{1}{2} \bar{\gamma}^{kl} \hat{\mathcal{D}}_k \hat{\mathcal{D}}_l \bar{\gamma}_{ij} \nonumber \\
&+ \bar{\gamma}_{k(i}\hat{\mathcal{D}}_{j)}(\tilde{\Lambda}^k - 2 e^{4 \phi} Z^k) + \mathcal{D}_i Z_j + \mathcal{D}_j Z_i \nonumber \\
&+ \Delta\Gamma^k \Delta\Gamma_{(ij)k} \nonumber \\
&+ \bar{\gamma}^{kl} (2 \Delta\Gamma^m_{k(i}\Delta\Gamma_{j)ml} + \Delta\Gamma^m_{ik} \Delta\Gamma_{mjl}).
\end{align}
Combining the terms $\bar{\gamma}_{k(i}\hat{\mathcal{D}}_{j)}(- 2 e^{4 \phi} Z^k)$
and $\mathcal{D}_i Z_j + \mathcal{D}_j Z_i$ it can be seen that all partial derivatives $\partial_i Z_j$
in $\bar{R}^{Z4}_{ij}$ cancel out exactly, meaning that $R^{Z4}_{ij}$ reduces to
\begin{equation}
\bar{R}^{Z4}_{ij} = \bar{R}_{ij} -8 Z_{(i} \partial_{j)} \phi  + 2 \gamma_{k(i} (\Gamma^k_{j)l}-\hat{\Gamma}^k_{j)l}) Z^l.
\end{equation}
With the above, and defining $\partial_0 \equiv \partial_t - \mathcal{L}_{\beta}$,
where $\mathcal{L}_{\beta}$ is the Lie derivative along the shift $\beta^i$,
we arrive at the following set of coordinate basis evolution equations for the fCCZ4 system:
\begin{empheq}[box=\exactfbox]{align*}
\partial_0 \bar{\gamma}_{ij} &= -\frac{2}{3}\bar{\gamma}_{ij}\bar{\mathcal{D}}_k\beta^k
- 2 \alpha \bar{A}_{ij}, \\ 
\partial_0 \phi &= \frac{1}{6}\bar{\mathcal{D}}_i \beta^i - \frac{1}{6} \alpha K, \\
\partial_0 \bar{A}_{ij} &= -\frac{2}{3} \bar{A}_{ij}\bar{\mathcal{D}}_k \beta^k
- 2 \alpha \bar{A}_{ik}\bar{A}^k_j
+\alpha \bar{A}_{ij} (K - 2\Theta) \nonumber \\
&+ e^{-4\phi}[-2\alpha \bar{\mathcal{D}}_i \bar{\mathcal{D}}_j \phi
+ 4 \alpha \bar{\mathcal{D}}_i\phi \bar{\mathcal{D}}_j \phi \nonumber \\
&+ 4 \bar{\mathcal{D}}_{(i}\alpha \bar{\mathcal{D}}_{j)}\phi
-\bar{\mathcal{D}}_i\bar{\mathcal{D}}_j \alpha \nonumber \\
&+ \alpha (\bar{R}^{Z4}_{ij} - 8\pi S_{ij})]^{\mathrm{TF}}, \\
\partial_0 K &= e^{-4\phi}[\alpha(\bar{R}^{Z4} - 8 \bar{\mathcal{D}}^i \bar{\mathcal{D}}_i \phi
- 8 \bar{\mathcal{D}}^2 \phi ) \nonumber \\
&- (2 \bar{\mathcal{D}}^i \alpha \bar{\mathcal{D}}_i \phi
+ \bar{\mathcal{D}}^2 \alpha)] + \alpha(K^2-2\Theta K) \nonumber \\
&- 3\kappa_1(1+\kappa_2)\Theta + 4\pi\alpha(S-3E),\\
\partial_0 \Theta & = \frac{1}{2} \alpha [e^{-4\phi}(\bar{R}^{Z4}
-8 \bar{\mathcal{D}}^i \phi \bar{\mathcal{D}}_i \phi
-8 \bar{\mathcal{D}}^2\phi) \nonumber \\
&- \bar{A}^{ij}\bar{A}_{ij} + \frac{2}{3}K^2 - 2\Theta K] \nonumber \\
&-Z^i\partial_i \alpha - \kappa_1 (2+\kappa_2)\Theta - 8\pi \alpha E, \\
\partial_0 \tilde{\Lambda}^i &= \bar{\gamma}^{jk}\hat{\mathcal{D}}_j\hat{\mathcal{D}}_k\beta^i
+ \frac{2}{3}\Delta\Gamma^i\bar{\mathcal{D}}_j \beta^j
+ \frac{1}{3}\bar{\mathcal{D}}^i \bar{\mathcal{D}}_j\beta^j \nonumber \\
&- 2 \bar{A}^{jk}(\delta^i_j \partial_k \alpha - 6 \alpha \delta^i_j \partial_k \phi
- \alpha \Delta \Gamma^i_{jk}) \nonumber \\
&- \frac{4}{3} \alpha \bar{\gamma}^{ij}\partial_j K
+ 2 \bar{\gamma}^{ij}(\alpha \partial_j \Theta - \Theta \partial_j \alpha)  -\frac{4}{3} \alpha K e^{4 \phi} Z^i \nonumber \\
&- 2 \kappa_1 e^{4 \phi} Z^i + 2 \kappa_3 e^{4 \phi}(\frac{2}{3} Z^i \hat{\mathcal{D}}_k\beta^k
- Z^k \hat{\mathcal{D}}_k \beta^i) \nonumber \\
&- 16 \pi \alpha \bar{\gamma}^{ij} S_j.
\end{empheq}
Here $[]^{\mathrm{TF}}$ denotes the trace-free part of a tensor:
$T^{{\mathrm{TF}}}_{ij} \equiv T_{ij} - \frac{1}{3} \gamma_{ij} T^k_{~k}$,
and $\kappa_3$ is a constant that determines the covariance of the equations,
in particular $\kappa_3=1$ (the choice adopted for all simulations presented
in this work), corresponds to full covariance~\cite{Alic2012}.
Unlike in~\cite{fCCZ42014}, we have absorbed all covariant derivatives
of $Z^i$ in $\bar{R}^{Z4}_{ij}$, meaning that no derivatives of the
constraint vector appear in the above equations.

During the time evolution, we continuously enforce
$\partial_t \bar{\gamma} =0$, as well as the constraint $\bar{A}^i_{~i} = 0$.
We have also implemented the $\chi=e^{4 \phi}$~\cite{Campanelli2006}
and $W=e^{2 \phi}$~\cite{Marronetti2008} variants
of the conformal factor evolution, resulting in the following evolution equations:
\begin{align}
\partial_0 \chi &= -\frac{2}{3} \chi \bar{\mathcal{D}}_i \beta^i
+ \frac{2}{3} \chi \alpha K, \\
\partial_0 W &= -\frac{1}{3} W \bar{\mathcal{D}}_i \beta^i
+ \frac{1}{3} W \alpha K.
\end{align}
The choice between these variants can be made at runtime.

In order to close the evolution system we need to choose a gauge.  For all
dynamical spacetime evolution simulations presented in this paper
we adopt the \enquote{standard gauge}, or \enquote{moving-puncture gauge}, meaning 1+log slicing
\begin{equation}
\partial_t \alpha = -2 (K- 2 \Theta) + \beta^i \partial_i \alpha
\end{equation}
for the lapse (see \cite{Bona1995}) and a $\Gamma$-driver\footnote{The $\Gamma$-driver owes its name to the
appearance of the conformal connection functions $\tilde{\Gamma}^i$
introduced in~\cite{Baumgarte1999}. While we use $\tilde{\Lambda}^i$ here, rather
than $\tilde{\Gamma}^i$, we still use the name $\Gamma$-driver for this gauge condition.}
\begin{eqnarray}
\partial_t \beta^i &=& B^i + \beta^j \hat{\mathcal{D}}_j \beta^i,\\
\partial_t B^i &=& \frac{3}{4} (\partial_t \tilde{\Lambda}^i - \beta^j \hat{\mathcal{D}}_j \tilde{\Lambda}^i)
+ \beta^j \hat{\mathcal{D}}_j B^i -\eta B^i
\end{eqnarray}
for the shift (see~\cite{Alcubierre2003}). Here $\eta$ is a damping parameter with
dimensions of inverse length, and we have adopted the covariant form of \cite{BrownCovariant2009}.
In the code, the inclusion of shift advection terms can be turned off by the user at runtime.

As noted above, the actual evolved tensors in the code are not in the
coordinate basis $\left\{\bar{\gamma}_{ij}, \bar{A}_{ij},
\tilde{\Lambda}^{i}, \beta^{i},
B^{i}=\dot{\beta}^{i}\right\}$, but rather the orthonormal
basis $\left\{h_{\{i\}\{j\}}, \bar{A}_{\{i\}\{j\}},
\tilde{\Lambda}^{\{i\}}, \beta^{\{i\}},
B^{\{i\}}=\dot{\beta}^{\{i\}}\right\}$, respectively. The remaining
evolved quantities $\left\{\alpha,K,\Theta,\phi\right\}$ are scalars and thus do not depend on choice of basis.

Finally, the matter sources in the spacetime evolution equations are given by projections of the stress-energy
tensor, namely
\begin{align}
E &\equiv n^{\mu} n^{\nu} T_{\mu \nu} = \frac{1}{\alpha^2} \left(T_{tt} - 2 \beta^i T_{t i} + \beta^i \beta^j T_{ij} \right),\\
S_i &\equiv - \gamma_{i \mu} n_{\nu} T^{\mu \nu} = -\frac{1}{\alpha} T_{ti} + \frac{1}{\alpha} \beta^j T_{ij}, \\
S_{ij} &\equiv \gamma_{i \mu} \gamma_{j \nu} T^{\mu \nu}, \\
S &\equiv \gamma^{ij} S_{ij}.
\end{align}
While we have only presented the fCCZ4 equations above, the user can select
to evolve the BSSN system at runtime as well.

%
\subsection{GRMHD in the reference-metric formalism}
\label{subsec:grmhd}
%
In this section we review the reference-metric formalism for GRHD presented
in~\cite{Montero2014,Baumgarte2015} and extend it to GRMHD, using a
vector potential evolution scheme to guarantee the absence of magnetic monopoles.

%
\subsubsection{Conservation laws in four-dimensional form}
%

The evolution of a magnetized fluid is governed by the conservation of baryon number
\begin{equation}\label{eq:4D_continuity}
\nabla_{\mu} (\rho u^{\mu}) = 0,
\end{equation}
which results in the continuity equation, and the conservation of energy-momentum
\begin{equation}\label{eq:4D_stress_energy}
\nabla_{\mu} T^{\mu \nu} =
\nabla_{\mu} (T^{\mu \nu}_{\mathrm{matter}} + T^{\mu \nu}_{\mathrm{EM}}) = 0,
\end{equation}
which results in the (relativistic) Euler equation and the conservation of total energy.
Assuming a perfect fluid, the fluid stress-energy tensor $T^{\mu \nu}_{\mathrm{matter}}$ is given by
\begin{equation}
T^{\mu \nu}_{\mathrm{matter}} = \rho h u^{\mu} u^{\nu} + P g^{\mu \nu},
\end{equation}
where $\rho$ is the rest mass density, $P$ the fluid pressure,
$h = 1+ \epsilon +P/\rho$ the specific enthalpy, $\epsilon$
the internal energy density, and $u^{\mu}$ the fluid four-velocity,
respectively.

In terms of the Faraday tensor $F^{\mu \nu}$, the EM stress-energy tensor is
\footnote{The Heaviside-Lorentz (HL) units we adopt in this work are rationalized,
as no explicit factors of $4 \pi$ appear in the Maxwell equations in these units.
Electric and magnetic fields in HL and Gauss units are therefore related by
a factor of $\sqrt{4 \pi}$: $E^{\mu}_{\mathrm{[Gauss]}} = \sqrt{4 \pi} E^{\mu}_{\mathrm{[HL]}}$,
$B^{\mu}_{\mathrm{[Gauss]}} = \sqrt{4 \pi} B^{\mu}_{\mathrm{[HL]}}$.
Consequently, in Gauss units, the EM stress-energy tensor is defined as
$T^{\mu \nu}_{\mathrm{EM}} = \frac{1}{4 \pi}(F^{\mu \lambda} F^{\nu}_{\lambda}
-\frac{1}{4} g^{\mu \nu} F^{\lambda \kappa} F_{\lambda \kappa})_{[\mathrm{Gauss}]}.$}
\begin{equation}
T^{\mu \nu}_{\mathrm{EM}} = F^{\mu \lambda} F^{\nu}_{\lambda}
-\frac{1}{4} g^{\mu \nu} F^{\lambda \kappa} F_{\lambda \kappa}.
\end{equation}
We next decompose the Faraday tensor $F^{\mu \nu}$ as
\begin{equation} \label{FaradayEB}
F^{\mu \nu} = U^{\mu} E^{\nu}_{(U)} - U^{\nu}E^{\mu}_{(U)}
+ \epsilon^{\mu \nu \lambda \kappa} U_{\lambda} B_{\kappa (U)},
\end{equation}
where
\begin{equation}
\epsilon^{\mu \nu \lambda \kappa} \equiv \frac{-1}{\sqrt{-g}}[\mu \nu \lambda \kappa],
\end{equation}
and where $[\mu \nu \lambda \kappa]$ is the totally antisymmetric Levi-Civita symbol
($=+(-)1$ for even (odd) permutations of $[0123]$, and 0 if any two indices are
repeated). Here $E^{\mu}_{(U)}$ and $B^{\mu}_{(U)}$ are the electric and magnetic
fields measured by an observer with {\it generic} four-velocity $U^{\mu}$,
\begin{equation}
E^{\mu}_{(U)} = F^{\mu\nu} U_\nu, \qquad
B^{\mu}_{(U)} = \frac{1}{2} \epsilon^{\mu \nu \kappa \rho} U_\nu F_{\rho \kappa}.
\end{equation}
Both $E^{\mu}_{(U)}$ and $B^{\mu}_{(U)}$ are orthogonal to $U^{\mu}$,
i.e.~$E^{\mu}_{(U)}U_{\mu} = B^{\mu}_{(U)} U_{\mu} = 0$.

In the following we focus on two observers of particular interest, namely
observers comoving with the fluid (i.e., with four-velocity
$U^\mu=u^\mu$) and normal observers (with four-velocity $U^\mu=n^\mu$).  Following convention,
we denote the fields observed by the former with $E^{\mu}_{(u)}$ and
$B^{\mu}_{(u)}$, but the latter simply with $E^\mu = E^\mu_{(n)}$ and $B^\mu = B^\mu_{(n)}$.

In the ideal MHD limit we assume that the fluid acts as a perfect conductor, meaning that
the electric field observed by an observer comoving with the fluid vanishes:
\begin{equation} \label{E_u_zero}
E^\mu_{(u)} = F^{\mu \nu} u_\nu = 0.
\end{equation}
Thus in this approximation $F^{\mu \nu}$, which generally depends on both
electric and magnetic fields,
can be expressed in terms of magnetic fields alone,
\begin{equation}
F^{\mu \nu} = \epsilon^{\mu \nu \lambda \kappa} u_{\lambda} B_{\kappa}^{(u)},
\end{equation}
and $^{*}F^{\mu \nu}$, the dual of the Faraday tensor, as
\begin{equation} \label{Faraday_dual_n}
^{*}F^{\mu \nu} = \frac{1}{2} \epsilon^{\mu \nu \lambda \kappa} F_{\lambda \kappa}
= u^\mu B_{(u)}^\nu - u^\nu B_{(u)}^\mu.
\end{equation}
In the ideal MHD limit we can also write the magnetic field $B^\mu_{(u)}$ as a projection of
$B^\mu$ along the fluid four-velocity $u^{\nu}$,
\begin{equation} \label{projection}
B^\mu_{(u)} = \frac{1}{W} P^\mu_{~\nu} B^\nu,
\end{equation}
where $P^\mu_{~\nu} =  g^\mu_{~\nu} + u^{\mu} u_{\nu}$, and where we have introduced
the Lorentz factor $W$ between the fluid and normal observers,
$W \equiv - n_\mu u^\mu = \alpha u^t$.
Inserting \eqref{projection} into \eqref{Faraday_dual_n} yields
\begin{equation} \label{Faraday_dual}
^{*}F^{\mu \nu} = \frac{1}{W} \left( u^{\mu} B^{\nu} - u^{\nu} B^{\mu} \right).
\end{equation}
Introducing the abbreviation\footnote{Similarly to differences in the
EM stress-energy tensor, treatments that adopt Gauss units (\cite{Villiers2003,
IllinoisGRMHD2005,Cosmos++2005,BonazzolaGRMHD2007,IGM2015})
rather than Lorentz-Heaviside units (\cite{Koide1999,HARM2003,Komissarov2004,SACRA2005,
ValenciaGRMHD2006,HAD2006,NobleHARM2006,
Whisky2007,CerdaDuran:2008pv,NobleHarm2009,X-ECHO2011,WhiskyGRRMHD2013,
GRHydro2014,Athena++2016,SpECTRE2017,BHAC2017,Spec_mag2018,FambriAder2018,Cipolletta:2019geh,H-AMR2019})
often use the definition
$b^{\mu}\equiv B^\mu_{(u)} / \sqrt{4 \pi}$ instead of \eqref{b}.  The resulting
fields $b^{\mu}$, however, are again identical in both treatments, so that
expressions for the stress-energy tensor, for example, take the same from
when written in terms of $b^{\mu}$, see \eqref{stress_energy_b}.}
\begin{equation} \label{b}
b^{\mu}\equiv B^\mu_{(u)}
\end{equation}
we can write the electromagnetic stress-energy tensor as
\begin{equation} \label{stress_energy_b}
T^{\mu \nu}_{\mathrm{EM}} = \left(u^{\mu} u^{\nu}
+\frac{1}{2} g^{\mu \nu} \right) b^2 - b^{\mu} b^{\nu},
\end{equation}
where $b^2\equiv b^{\mu}b_{\mu}$; so that the total stress-energy tensor $T^{\mu \nu}$ becomes
\begin{equation}\label{eq::tmunu_perfect_fluid}
T^{\mu \nu} = \rho\, h^{*} u^{\mu} u^{\nu} + P^{*} g^{\mu \nu} - b^{\mu} b^{\nu}.
\end{equation}
Here we have defined $h^{*}=1+\epsilon+(P+b^2)/\rho$ as the magnetically modified specific enthalpy
and $P^{*}=P+b^2/2$ as the magnetically modified isotropic pressure.
Finally, the evolution of the magnetic field is governed by the
homogeneous Maxwell equations
\begin{equation}\label{eq:4D_maxwell}
\nabla_{\nu} (^{*}F^{\mu \nu}) = 0.
\end{equation}

\subsubsection{The 3+1 GRMHD equations in the reference-metric formalism}
%

We now recast the above conservation laws using both a 3+1 split and a reference-metric approach.
The result will be a set of equations that is suitable for
numerical integration in spherical coordinates, and that meshes well with the form of the field equations
as presented in Sec.~\ref{subsec:fCCZ4}.  The fluid equations have been previously derived in
\cite{Montero2014,Baumgarte2015}, and we will extend the formalism to the GRMHD equations.

The key idea is to repeatedly use identities for divergences (see, e.g., Problem 7.7 in \cite{Lightman1975}).
For the continuity equation \eqref{eq:4D_continuity}, for example, we use
\begin{equation}\label{eq:cov_deriv_vector}
\nabla_{\nu} V^{\nu} = \frac{1}{\sqrt{|g|}} \partial_{\nu} (\sqrt{|g|} V^{\nu})
\end{equation}
to arrive at
\begin{align} \label{continuity_valencia}
0 &= \nabla_{\nu} (\rho u^{\nu}) = \frac{1}{\sqrt{-g}} \partial_{\nu}(\sqrt{-g} \rho u^{\nu}) \nonumber \\
&= \frac{1}{\sqrt{-g}} \left(\partial_t(\sqrt{-g} \rho u^t) +\partial_i(\sqrt{-g} \rho u^i) \right) \nonumber \\
&= \partial_t (e^{6 \phi} \sqrt{\bar{\gamma}} \rho W)
+ \partial_i \left(\alpha e^{6 \phi} \sqrt{\bar{\gamma}} \rho W \bar{v}^i \right).
\end{align}
Here we have defined
\begin{equation}
v^i \equiv \frac{1}{W} \gamma^i_{~\nu} u^\nu = \frac{u^i}{W}+\frac{\beta^i}{\alpha}
\end{equation}
as the fluid three-velocity and
\begin{equation}
\bar{v}^i \equiv v^i-\frac{\beta^i}{\alpha} = \frac{u^i}{W}
\end{equation}
as the~\enquote{advection velocity}.  We have also written
the square root of the determinant of the spacetime metric
$\sqrt{-g}$ as
\begin{equation}
\sqrt{-g} = \alpha \sqrt{\gamma} = \alpha e^{6 \phi} \sqrt{\bar{\gamma}}.
\end{equation}

Equation \eqref{continuity_valencia} is the continuity equation in a form that is often referred to as the
\enquote{Valencia} form of the equations (see~\cite{Banyuls1997}).  This version of the equations is well suited
for simulations in Cartesian coordinates, but in curvilinear coordinates the vanishing of the determinant
$\bar \gamma$ may cause numerical problems.  Following \cite{Montero2014}, we apply the
identity \eqref{eq:cov_deriv_vector} again, and convert the partial derivatives $\partial_i$ to
covariant derivatives $\hat{\mathcal{D}}_i$ associated with the
background (reference) metric
$\hat{\gamma}_{ij}$,\footnote{We will assume throughout that the reference-metric $\hat \gamma_{ij}$
is independent of time.}
\begin{equation}
0 = \partial_t (e^{6 \phi} \sqrt{\bar{\gamma}/\hat{\gamma}}  \rho W)
+ \hat{\mathcal{D}}_i (\alpha e^{6 \phi} \sqrt{\bar{\gamma}/\hat{\gamma}}\, \rho W \bar{v}^i).
\end{equation}
Note that the combination $\bar \gamma / \hat \gamma$ remains finite for regular spacetimes.
We now define the conserved density $D$
\begin{equation}
D \equiv e^{6 \phi} \sqrt{\bar{\gamma}/\hat{\gamma}}  \rho W
\end{equation}
and the conserved density flux $f_D$
\begin{equation}
(f_{D})^i \equiv \alpha D \, \bar{v}^i
\end{equation}
to write the continuity equation in the form
\begin{equation}\label{eq:continuity_Dhat}
\partial_t D + \hat {\mathcal D}_i (f_{D})^i = 0.
\end{equation}
We note that a similar strategy is followed in the {\tt CoCoNuT} code,
without explicitly mentioning the reference-metric (see e.g.~\cite{dimmelmeierPhD2001}).
For reasons that will become apparent in Sec.~~\ref{subsec:integral_form} below,
we will implement the equation numerically as:
\begin{equation}\label{eq:continuity_christoffel}
\partial_t D + \partial_i (f_{D})^i = -(f_{D})^i \hat{\Gamma}^j_{ij}.
\end{equation}

We proceed similarly for the conservation of stress energy \eqref{eq:4D_stress_energy},
except that we now use the identity
\begin{equation}
\nabla_{\lambda} T^{\lambda}_{\mu}
= \frac{1}{\sqrt{|g|}} \partial_{\lambda} (\sqrt{|g|} T^{\lambda}_{\mu})
- T^{\lambda}_{\sigma} \Gamma^{\sigma}_{\lambda \mu}
\end{equation}
twice.  The spatial projection of \eqref{eq:4D_stress_energy} then yields the
relativistic Euler equation
\begin{equation}
\partial_t S_j + \hat {\mathcal D}_i (f_{S})_j^i
= (s_{S})_j
\end{equation}
or
\begin{equation} \label{Euler}
\partial_t S_j + \partial_i (f_{S})_j^i
= (s_{S})_j - (f_{S})_j^i \hat{\Gamma}^k_{ik}
+ (f_{S})^i_k \hat{\Gamma}^k_{ij},
\end{equation}
where we have defined the conserved momentum
\begin{equation}
S_j \equiv e^{6 \phi} \sqrt{\bar{\gamma}/\hat{\gamma}} T^i_j
=e^{6 \phi} \sqrt{\bar{\gamma}/\hat{\gamma}}  (\rho h^{*} W^2 v_j - \alpha b^0 b_j),
\end{equation}
the conserved momentum density fluxes
\begin{equation}
(f_{S})_j^i \equiv \alpha (S_j \bar{v}^i
+ e^{6 \phi}\sqrt{\bar{\gamma}/\hat{\gamma}}\,P^{*}\delta^i_j
- e^{6 \phi}\sqrt{\bar{\gamma}/\hat{\gamma}}\, b_j B^i/W),
\end{equation}
and where the source term is given by
\begin{align}
(s_{S})_j &= \alpha e^{6 \phi} \sqrt{\bar{\gamma}/\hat{\gamma}} \left( -T^{00} \alpha \partial_j \alpha
+ T^0_{i} \hat{\mathcal{D}}_j \beta^i  \right. \nonumber \\
&\left. + \frac{1}{2} (T^{00} \beta^i \beta^k + 2 T^{0 i} \beta^k
+ T^{ik}) \hat{\mathcal{D}}_j \gamma_{ik} \vphantom{\int_1^2} \right)
\end{align}
(see~\cite{Montero2014} for a detailed derivation).  The terms
$\hat{\mathcal{D}}_j \gamma_{ik}$ can be evaluated from
\begin{equation}
\hat{\mathcal{D}}_j \gamma_{ik} = e^{4 \phi} (4 \bar{\gamma}_{ik} \partial_j \phi
+ \hat{\mathcal{D}}_j \bar{\gamma}_{ik}),
\end{equation}
where the $\hat{\mathcal{D}}_j \bar{\gamma}_{ik} = \hat{\mathcal{D}}_j h_{ik}$
are computed already in \eqref{Delta_Gamma}.

Projecting the conservation of stress energy~\eqref{eq:4D_stress_energy}
along $n_{\nu}$ and subtracting the continuity Eq.~\eqref{eq:4D_continuity}
yields
\begin{equation}\label{eq:4d_energy_equation}
\nabla_{\mu} \left(n_{\nu}T^{\nu \mu} - \rho u^{\mu} \right)
=  T^{\mu \nu} \nabla_{\nu}n_{\mu}.
\end{equation}
We again apply the identity \eqref{eq:cov_deriv_vector} twice to arrive at the energy equation
\begin{equation}\label{eq:3d_energy_equation}
\partial_t \tau + \hat {\mathcal D}_i (f_{\tau})^i = s_{\tau}
\end{equation}
or
\begin{equation}
\partial_t \tau + \partial_i (f_{\tau})^i = s_{\tau} - (f_{\tau})^i \hat{\Gamma}^j_{ij},
\end{equation}
where we have defined $\tau$ as the total conserved energy
density subtracting the conserved density $D$\footnote{The motivation
to subtract the continuity equation from the projection
of the stress-energy conservation along $n_{\mu}$\eqref{eq:4d_energy_equation}
was to arrive at an evolution for $\tau$: It correctly recovers the Newtonian
limit and is numerically more accurate than evolving the total conserved energy
density (see e.g.~\cite{ValenciaGRMHD2006,Baumgarte2010,rezzolla2013relativistic}).}:
\begin{equation}
\tau \equiv e^{6 \phi} \sqrt{\bar{\gamma}/\hat{\gamma}}  (\rho h^{*} W^2 - P^{*} - (\alpha b^0)^2)-D,
\end{equation}
the conserved energy flux
\begin{equation}
(f_{\tau})^i \equiv \alpha (\tau \bar{v}^i + e^{6 \phi}\sqrt{\bar{\gamma}/\hat{\gamma}}\,P^{*}v^i
- \alpha e^{6 \phi}\sqrt{\bar{\gamma}/\hat{\gamma}}\, b^0 B^i/W),
\end{equation}
and where the source term is given by~\cite{Montero2014}
\begin{align}
s_{\tau} &= \alpha e^{6 \phi} \sqrt{\bar{\gamma}/\hat{\gamma}} (T^{00}(\beta^i \beta^j K_{ij}
- \beta^i \partial_i \alpha) \nonumber  \\
&+ T^{0i} (2 \beta^j K_{ij} - \partial_i \alpha) + T^{ij} K_{ij}).
\end{align}

In the above equations, both $B^i$ and $b^\mu = B^\mu_{(u)}$
make an appearance. The two fields are related by \eqref{projection}, so that we can always
compute $B^\mu_{(u)}$ from $B^\mu$.   Specifically, we contract \eqref{projection} with $n_\mu$ to obtain
\begin{equation}
b^0 = \frac{W B^i v_i}{\alpha},
\end{equation}
while a spatial projection of \eqref{projection} yields
\begin{equation}
b^i = \frac{B^i}{W} + W (B^j v_j) \bar v^i.
\end{equation}
We also have
\begin{equation}
b^2 = \frac{B^i B_i}{W^2} + (B^i v_i)^2.
\end{equation}

We now adopt the same approach to rewrite Maxwell's equations \eqref{eq:4D_maxwell}.
Since the Faraday tensor (as well as its dual) is antisymmetric, we now use the identity
\begin{equation} \label{identity_antisymmetric}
\nabla_{\nu} A^{\mu \nu} = \frac{1}{\sqrt{|g|}} \partial_{\nu} \left( \sqrt{|g|} A^{\mu \nu} \right),
\end{equation}
for antisymmetric tensors $A^{\mu \nu}$.  Inserting \eqref{Faraday_dual} into \eqref{eq:4D_maxwell}
and using \eqref{identity_antisymmetric} we obtain
\begin{equation} \label{maxwell_temp}
\partial_t (\sqrt{\gamma} B^\mu) = \partial_i \left( \frac{\alpha \sqrt{\gamma}}{W} (u^\mu B^i - u^i B^\mu) \right) .
\end{equation}
The temporal component of this equation results in the solenoidal constraint, stating the
absence of magnetic monopoles,
\begin{equation} \label{absence_of_monopoles_1}
\partial_i ( \sqrt{\gamma} B^i ) = \sqrt{\hat \gamma} \, \hat {\mathcal D}_i ( e^{6 \phi} \sqrt{\bar \gamma / \hat \gamma} B^i ) = 0.
\end{equation}
We now define
\begin{equation} \label{curly_B_defined}
\mathcal{B}^i \equiv e^{6 \phi} \sqrt{\bar{\gamma}/\hat{\gamma}}  B^i,
\end{equation}
so that \eqref{absence_of_monopoles_1} reduces to
\begin{equation} \label{absence_of_monopoles_2}
\hat {\mathcal D}_i {\mathcal B}^i = 0.
\end{equation}
For the spatial part of \eqref{maxwell_temp} we use \eqref{identity_antisymmetric} again to obtain
the induction equation in 3+1 form
\begin{equation} \label{eq:induction3+1}
\partial_t \mathcal{B}^j
= \hat{\mathcal{D}}_i \left(\alpha (\bar{v}^j \mathcal{B}^i - \bar{v}^i \mathcal{B}^j) \right).
\end{equation}
This form of the solenoidal constraint and continuity equation is
very similar to the one presented in~\cite{CerdaCuranFluxCT2007,CerdaDuran:2008pv},
but solved there using a constraint transport approach, while we
evolve the vector potential of the magnetic field in our framework instead.

%
\subsubsection{Vector potential evolution equations}
%

Numerically evolving the induction equation \eqref{eq:induction3+1} directly is generally
problematic, since accumulating numerical error will typically result in the magnetic field
having nonvanishing divergence.  The resulting growth of spurious magnetic monopoles has severe
consequences of the evolution, since it will result in nonphysical fluid acceleration
in the direction of the magnetic field (see, e.g., \cite{Brackbill1980}).

Various approaches have been implemented
to avoid this growth of magnetic monopoles in (GR)MHD simulations (see~\cite{TothDivB2000}
for a comprehensive overview). The three most commonly adopted approaches in GRMHD codes
are (i) hyperbolic divergence cleaning via a generalized Lagrange multiplier~\cite{Dedner2002}; (ii)
constrained transport (CT)~\cite{EvansCT1988} schemes in which the magnetic field is updated in
such a way that the divergence (measured in a finite-difference stencil that is
compatible with the base CT scheme) remains unchanged to round off during the evolution; and (iii)
evolving not the magnetic field directly but rather its vector potential and
taking the curl of the vector potential in order to compute the magnetic
field~\cite{IllinoisGRMHD2005,Whisky2007,Cipolletta:2019geh}.
As the divergence of the curl of a vector field is identically zero,
the latter approach guarantees a solenoidal magnetic field to
round off error during the evolution.

In developing the GRMHD evolution framework in spherical coordinates,
we opted to implement the latter, namely evolving the
vector potential in a cell-centered fashion.
We choose a vector-potential formulation for four reasons: (1) the
resulting equations can be easily incorporated into our
reference-metric formalism; (2) contrary to the hyperbolic
divergence cleaning, the solenoidal constraint is automatically fulfilled to machine
precision; (3) there is no need to extend our internal parity boundary conditions
to deal with the staggered magnetic fields used in CT schemes (though
see the {\tt CoCoNuT}~\cite{CerdaCuranFluxCT2007,CerdaDuran:2008pv} and {\tt Aenus}~\cite{Obergaulinger2008}
codes for implementations of staggered CT schemes in spherical
coordinates); and (4) it has recently been
shown to be strongly hyperbolic~\cite{HilditchHyperbolicity2019}.

Imposing the ideal MHD limit again,
and taking a projection of Eq.~\eqref{E_u_zero} with the spatial metric \eqref{spatial_metric} shows that
the electric and magnetic field as observed by the normal observer are related by
\begin{equation} \label{E_mu_1}
E^{\mu} = \frac{-1}{W} \epsilon^{\mu \nu \lambda \kappa} u_{\nu} n_{\lambda} B_{\kappa}.
\end{equation}
Defining the three-dimensional antisymmetric tensor as
\begin{equation}
\epsilon_{\mu \nu \lambda} = n^{\kappa} \epsilon_{\kappa \mu \nu \lambda}, \mathrm{~~or~~}
\epsilon^{\mu \nu \lambda} = n_{\kappa} \epsilon^{\kappa \mu \nu \lambda},
\end{equation}
so that $\epsilon_{ijk} = \sqrt{\gamma} \, [ijk]$,
we may rewrite \eqref{E_mu_1} as
\begin{equation}  \label{E_MHD}
E_i = -\epsilon_{ijk} \bar{v}^j B^k,
\end{equation}
where we have used $\epsilon_{tij} = - \beta^k \epsilon_{ikj}$ (see Eq.~(32) in~\cite{BaumgarteGRMHD2003}).

We now introduce a four-vector potential
\begin{equation}
\mathcal{A}_{\mu} = \Phi n_{\mu} + A_{\mu},
\end{equation}
where $\Phi$ is the electromagnetic scalar potential and $A_\mu$ is purely spatial, $A_{\mu}n^{\mu}=0$,
so that
$A_t = \beta^i A_i$
and ${\mathcal A}_t = - \alpha \Phi + \beta^i A_i$.
Writing the Faraday tensor in terms of $\mathcal{A}_{\mu}$
yields
\begin{equation}\label{eq:faraday_in_terms_of_Amu}
F_{\mu \nu} = \partial_{\mu} \mathcal{A}_{\nu} - \partial_{\nu} \mathcal{A}_{\mu}
=n_{\mu} E_{\nu} + n_\nu E_\mu +  \epsilon_{\mu \nu \lambda} B^{\lambda}.
\end{equation}
Contracting this with $\epsilon^{\mu \nu \lambda}$ yields
\begin{equation}
\epsilon^{\mu \nu \lambda} (\partial_{\mu} \mathcal{A}_{\nu} - \partial_{\nu} \mathcal{A}_{\mu})
=\epsilon^{\mu \nu \lambda} \epsilon_{\mu \nu \kappa} B^{\kappa} = 2 B^\lambda,
\end{equation}
or
\begin{equation}
B^i = \epsilon^{ijk} \partial_j A_k.
\end{equation}
Inserting the definition \eqref{curly_B_defined} then results in
\begin{equation} \label{eq:curl_of_A}
{\mathcal B}^i = \hat \epsilon^{ijk} \partial_j A_k = \hat \epsilon^{ijk} \hat {\mathcal D}_j A_k,
\end{equation}
where $\hat{\epsilon}_{ijk} = \sqrt{\hat{\gamma}} \, [ijk]$ and $\hat{\epsilon}^{ijk} = \hat{\gamma}^{-1/2} \, [ijk]$.

Finally, we may evaluate a mixed time-space component of \eqref{eq:faraday_in_terms_of_Amu} to
find
\begin{equation}
\partial_t A_i = - \alpha E_i + \epsilon_{tij} B^j - \hat{\mathcal{D}}_i (\alpha \Phi - \beta^j A_j).
\end{equation}
Inserting \eqref{E_MHD} and expressing the result in terms of ${\mathcal B}^i$ results in
\begin{align}
\partial_t A_i
= \alpha \hat{\epsilon}_{ijk} \bar{v}^j \mathcal{B}^k -\hat{\mathcal{D}}_i \left(\alpha \Phi -\beta^j A_j \right).
\end{align}
In our code, we evolve the spatial components of the vector potential $A_i$, and compute, at each
time step, the conserved magnetic field from \eqref{eq:curl_of_A}.

We evolve the vector potential in a \enquote{generalized Lorenz gauge}~\cite{Farris2012PRL}
\begin{equation}
\nabla_{\mu} A^{\mu} = - \zeta \Phi,
\end{equation}
where $\zeta$ is a damping parameter with dimensions of inverse length.  As before, we use the identity
\eqref{eq:cov_deriv_vector} twice to rewrite this as
\begin{align} \label{Phi_evolve}
\partial_t (e^{6 \phi} \sqrt{\bar{\gamma} / \hat \gamma} \,\Phi) &
+ \hat {\mathcal D}_i (\alpha e^{6 \phi} \sqrt{\bar{\gamma} / \hat \gamma} \, A^i
- e^{6 \phi} \sqrt{\bar{\gamma} / \hat \gamma} \beta^i \Phi) \nonumber \\
&= -\zeta \alpha e^{6 \phi} \sqrt{\bar{\gamma} / \hat \gamma} \Phi.
\end{align}
We now define
\begin{equation}
\hat \Phi \equiv e^{6 \phi} \sqrt{\bar{\gamma} / \hat \gamma} \Phi
\end{equation}
and
\begin{equation}
(f_{\Phi})^i \equiv \alpha e^{6 \phi} \sqrt{\bar{\gamma}/\hat{\gamma}}\, A^i
- \beta^i \hat \Phi.
\end{equation}
and evolve \eqref{Phi_evolve} as
\begin{align}
\partial_t \hat \Phi + \partial_i (f_{\Phi})^i
= -\zeta \alpha \hat \Phi - (f_{\Phi})^i \hat{\Gamma}^j_{ij}.
\end{align}
In all applications shown in this paper we followed~\cite{IGM2015} and
chose $\zeta = 1.5 / \Delta t$,\footnote{Like the $\Gamma$-driving
  shift parameter $\eta$, $\zeta$ has
units of $1/t$ (or equivalently 1/$M$ in $G=c=1$ units). Thus the value of $\zeta$ is constrained by
the Courant–Friedrichs–Lewy (CFL) condition (in precisely the same way as described
by~\cite{Schnetter2010} for $\eta$). However, unlike $\eta$, we prefer the
damping provided by $\zeta>0$ to be as strong as possible
everywhere. Our choice $\zeta = 1.5/\Delta t$ is quite strong, but
should be stable for CFL factors of $\approx \frac{2}{3}$ or smaller,
consistent with the required CFL factors of $\frac{1}{2}$ or smaller
when solving the BSSN/CCZ4 equations. As a corollary, when convergence
testing $\zeta=1.5/\Delta t$ must correspond to the lowest-resolution
simulation's $\Delta t$, to ensure the CFL condition is not violated.}
where $\Delta t$ is the global time step of our numerical evolution.

We note that our algorithm is not staggered, i.e.~the vector potential $A_i$ lives at the
cell centers as do all other variables. In order to update the magnetic field, we calculate the
curl of $A_i$~\eqref{eq:curl_of_A}, where we apply the product rule and
take derivatives of the scale factors analytically, as with all other fields.  Initial data for
magnetic fields are generated in the same way, namely by taking the curl of a
prescribed initial vector potential $A_i$.

%
\subsubsection{Summary}
\label{sec:summary}
%

In summary, the GRMHD evolution system in the reference-metric formalism is
composed of the following conserved quantities:
\begin{align}
D &\equiv e^{6 \phi} \sqrt{\bar{\gamma}/\hat{\gamma}}  \rho W, \\
S_j 			&\equiv e^{6 \phi} \sqrt{\bar{\gamma}/\hat{\gamma}}  (\rho h^{*} W^2 v_j - \alpha b^0 b_j), \\
\tau 			&\equiv e^{6 \phi} \sqrt{\bar{\gamma}/\hat{\gamma}}  (\rho h^{*} W^2 - P^{*} - (\alpha b^0)^2)-D, \\
\mathcal{B}^j 	&\equiv e^{6 \phi} \sqrt{\bar{\gamma}/\hat{\gamma}}  B^j,
\end{align}
where we note that, unlike the corresponding conserved variables of the
{\it Valencia formulation}, these are true scalars and vectors (tensor densities
of weight zero). The GRMHD evolution
equations in coordinate basis are:
\begin{empheq}[box=\exactfbox]{align*}
\partial_t D + \partial_i (f_{D})^i
&= -(f_{D})^i \hat{\Gamma}^j_{ij}, \nonumber \\
\partial_t S_j + \partial_i (f_{S})_j^i
&= (s_{S})_j - (f_{S})_j^i \hat{\Gamma}^k_{ik}
+ (f_{S})^i_k \hat{\Gamma}^k_{ij}, \nonumber \\
\partial_t \tau + \partial_i (f_{\tau})^i
&= s_{\tau} - (f_{\tau})^i \hat{\Gamma}^j_{ij}, \nonumber \\
\partial_t A_i &= \alpha \hat{\epsilon}_{ijk} \bar{v}^j \mathcal{B}^k
-\hat{\mathcal{D}}_i \left(\alpha \Phi -\beta^j A_j \right), \nonumber \\
\partial_t \hat \Phi + \partial_i (f_{\Phi})^i
&= -\zeta \alpha \hat \Phi - (f_{\Phi})^i \hat{\Gamma}^j_{ij}, \nonumber \\
\mathcal{B}^i &= \hat{\epsilon}^{ijk} \partial_j A_k. \nonumber
\end{empheq}

Before proceeding with the description of the GRMHD evolution equations expressed
in the orthonormal basis with respect to the spherical background metric, we note
that the geometric source terms introduced by rewriting the
equations in the reference-metric formalism break the roundoff level
conservation of baryon number when evolving $D$ in a finite volume scheme.
This is due to the fact that the resulting finite volume scheme is not
well-balanced~\cite{GreenbergWellBalanced1996} unless the geometric source terms are evaluated
in such a way as to numerically exactly cancel the fluxes through the cell surface
(see~\cite{FambriAder2018} for developments towards well-balanced schemes in GRMHD;
an extension that is beyond the scope of this work). This nonconservation is a drawback
of the scheme. For the conserved momenta $S_i$ and conserved energy $\tau$ this problem
is less severe, as those quantities are only strictly conserved in the presence of spacetime
symmetries~\cite{PapadopoulosCons1999}, due to the appearance of (spacetime)-geometric
source terms.
In a sense, both problems are similar: both nonconservations arise from rewriting covariant
derivatives in terms of partial derivatives which are suitable for the numerical integration
of the resulting evolution equations.

When numerically evolving the Euler equation in spherical coordinates,
there is nonconservation of momentum in the $\theta$ coordinate, which is due to
the presence of the \enquote{naked pressure term} $e^{6 \phi} \sqrt{\bar{\gamma}} P^{*}\delta^a_j$
which causes the breaking of zero-force equilibria. This is due to the $\theta$ dependence of the
spherical background metric introduces a pressure gradient even in the absence of forces,
as the finite volume scheme is not well balanced. One key advantage
of our reference-metric approach is its automatic conservation of $\theta$-momentum
due to the absence of the reference-metric determinant.
We note that there are various other strategies to deal with this problem in spherical
coordinates, see e.g.~\cite{Neilsen2000,CallAdvection2010,McKinney2012}.
The $\varphi$ coordinate is not affected by this, and angular momentum is therefore
identically conserved in spherical coordinates.

%
\subsection{Equations in orthonormal basis of spherical background metric}
\label{subsec:rescaling}
%
Before we continue to describe the implementation of the evolution system
in a finite volume method in Sec.~\ref{subsec:integral_form} below, we first make the
following choice for the initial determinant of the conformally related metric,
\begin{equation}
\bar \gamma = \hat \gamma  \mbox{~~~~~~~~~~at~} t = 0.
\end{equation}
Moreover,
we adopt the \enquote{Lagrangian choice} \eqref{eq:lagrangian_choice},
$\partial_t \bar{\gamma} = 0$, meaning that we have $\bar \gamma = \hat \gamma$
at all times (as noted in Sec.~\ref{subsec:fCCZ4}, this is continuously
enforced in the spacetime evolution).  Accordingly, the ratio $\bar \gamma / \hat \gamma$ is unity
and our definitions of the GRMHD variables therefore reduce to
\begin{align}
D 				&= e^{6 \phi} \rho W, \\
S_j 			&= e^{6 \phi} (\rho h^{*} W^2 v_j - \alpha b^0 b_j), \\
\tau 			&= e^{6 \phi} (\rho h^{*} W^2 - P^{*} - (\alpha b^0)^2)-D, \\
\mathcal{B}^j 	&= e^{6 \phi} B^j.
\end{align}
Using the orthonormal basis with respect to the background metric defined
above~\eqref{eq:hatted_basis_vectors},
we write the continuity equation as
\begin{equation}
  \partial_t D + \partial_i \left( \hat{\mathbf{e}}^{i}_{\{j\}}
  ({f}_{D})^{\{j\}} \right)
  = -(f_{D})^i \hat{\Gamma}^j_{ij}. \label{eq:fl}
\end{equation}
Note that the vector $(f_D)$ is expressed in both bases in
Eq.~\eqref{eq:fl}.
We evaluate the flux divergence by using the product rule and analytically
differentiating the scale factors,
\begin{equation}
\partial_i \left( \hat{\mathbf{e}}^{i}_{\{j\}}
({f}_{D})^{\{j\}} \right)
= \left( \partial_i \hat{\mathbf{e}}^{i}_{\{j\}} \right) ({f}_{D})^{\{j\}}
+ \hat{\mathbf{e}}^{i}_{\{j\}} \partial_i ({f}_{D})^{\{j\}},
\end{equation}
so that the continuity equation is given by
\begin{equation}
\partial_t D + \hat{\mathbf{e}}^{i}_{\{j\}} \partial_i ({f}_{D})^{\{j\}}
= -(f_{D})^{i} \hat{\Gamma}^j_{ij}
- \left( \partial_i \hat{\mathbf{e}}^{i}_{\{j\}} \right) ({f}_{D})^{\{j\}}.
\end{equation}
The momentum equation similarly becomes
\begin{align}
&\partial_t \left(\hat{\mathbf{e}}^{\{i\}}_{j} {S}_{\{i\}} \right)
+ \hat{\mathbf{e}}^{\{k\}}_{j} \hat{\mathbf{e}}^{i}_{\{l\}}
\partial_i \left( ({f}_{S})_{\{k\}}^{\{l\}} \right) \nonumber \\
&= - \partial_i \left(\hat{\mathbf{e}}^{\{k\}}_{j} \hat{\mathbf{e}}^{i}_{\{l\}} \right)({f}_{S})_{\{k\}}^{\{l\}}
-(f_{S})_{j}^{i} \hat{\Gamma}^k_{ik} +(f_{S})^{i}_{k} \hat{\Gamma}^k_{ij} + (s_{S})_j.
\end{align}
Before we write out the rest of the GRMHD equations, we note that the diagonality of the spherical background metric allows us to simplify
the notation in expressions containing the basis elements. For instance, in expressions
such as $\partial_t \left(\hat{\mathbf{e}}^{\{i\}}_{j} \hat{S}_{\{i\}} \right)$,
$\hat{\mathbf{e}}^{\{i\}}_{j} = 0$ if $i \neq j$, which leads to element-wise
multiplication of vector and tensor components in the orthonormal basis and
the basis elements. We could therefore write
$\hat{\mathbf{e}}^{\{j\}}_{j} \hat{S}_{\{j\}}$ where the
summation convention does not apply on repeated coordinate $(j)$ and
orthonormal $(\{i\})$ indices, but this can lead to significant confusion when interpreting equations.
Instead, we introduce a new notation defining the following vectors
and matrices of rescale factors $\hat{\mathcal{R}}_{\{i\}}$
(corresponding to the scale factors $h_i$ of the reference-metric):
\begin{align}
  \hat{\mathcal{R}}_{\{i\}} &\equiv \begin{pmatrix}
1 \\
r \\
r \sin \theta
\end{pmatrix} \\
  \hat{\mathcal{R}}^{\{i\}} &= 1/\hat{\mathcal{R}}_{\{i\}}, \\
  \hat{\mathcal{R}}_{\{i\}\{j\}} &= \hat{\mathcal{R}}_{\{i\}} \hat{\mathcal{R}}_{\{j\}}, \\
  \hat{\mathcal{R}}^{\{i\}}_{\{j\}} &= \hat{\mathcal{R}}^{\{i\}} \hat{\mathcal{R}}_{\{j\}}.
\end{align}
As explained above, the components of the orthonormal basis form a diagonal matrix,
and the rescaling amounts to element-wise multiplication
of vector and tensor components with the corresponding vectors
and matrices of rescaling factors (essentially computing the
Hadamard product). In order to write this in a notation that
can be used with the summation convention that we use throughout,
we define the following symbols that will be used to
express coordinate vectors and tensors in the orthonormal basis
with respect to $\hat{\gamma}_{ij}$.
\begin{equation}
  \sigma^{i}_{\{j\}\{k\}} = \sigma^{\{j\}\{k\}}_{i} \equiv \begin{cases}
		1  & i=j=k \\
		0 & \mathrm{otherwise}
\end{cases}
\end{equation}
Note that $\sigma^{i}_{\{j\}\{k\}}$ and $\sigma^{\{j\}\{k\}}_{i}$ are {\em not} tensors and that the indices in  both are not raised and lowered by the metric. In addition, indices of the two $\sigma$ matrices can be contracted with both coordinate and orthonormal indices (i.e., indices surrounded by curly braces).
As an example of using this notation for vectors and tensors,
the coordinate three-velocity is written as
\begin{equation}
  v^i = \hat{\mathbf{e}}^{i}_{\{j\}} {v}^{\{j\}} = \sigma^i_{\{j\}\{k\}} \hat{\mathcal{R}}^{\{j\}} {v}^{\{k\}},
\end{equation}
and the coordinate conformally related metric as
\begin{align}
\bar{\gamma}_{ij} &= \hat{\mathbf{e}}^{\{k\}}_{i} \hat{\mathbf{e}}^{\{l\}}_{j}
\left(\delta_{\{k\}\{l\}} + h_{\{k\}\{l\}} \right) \nonumber \\
  &= \sigma^{\{k\}\{m\}}_i \sigma^{\{l\}\{n\}}_j \hat{\mathcal{R}}_{\{m\}} \hat{\mathcal{R}}_{\{n\}}
  (\delta_{\{k\}\{l\}} + h_{\{k\}\{l\}}).
\end{align}
In this notation, the continuity, momentum and energy equations are written as
\begin{align}
  \partial_t D &+ \sigma^i_{\{j\}\{k\}} \hat{\mathcal{R}}^{\{j\}} \partial_i ({f}_{D})^{\{k\}} \nonumber \\
  &= -(f_{D})^i \hat{\Gamma}^j_{ij} -\sigma^i_{\{j\}\{k\}} (\partial_i \hat{\mathcal{R}}^{\{j\}})({f}_{D})^{\{k\}}, \\
  \partial_t {S}_{\{j\}} &+\sigma^{i}_{\{k\}\{l\}} \delta^{n}_j
  \hat{\mathcal{R}}^{\{k\}} \partial_i \left(({f}_{S})_{\{n\}}^{\{l\}} \right) \nonumber \\
  &= \Big[- \sigma^{i}_{\{k\}\{l\}} \sigma^{\{m\}\{n\}}_q
  ({f}_{S})_{\{m\}}^{\{k\}} \left(\partial_i \hat{\mathcal{R}}^{\{l\}}_{\{n\}} \right) \nonumber \\
&- (f_{S})_q^i \hat{\Gamma}^k_{ik} + (f_{S})^i_k \hat{\Gamma}^k_{iq} + (s_{S})_q \Big] \sigma_{\{j\}\{p\}}^{q} \hat{\mathcal{R}}^{\{p\}}, \\
  \partial_t \tau &+ \sigma^i_{\{j\}\{k\}} \hat{\mathcal{R}}^{\{j\}} \partial_i ({f}_{\tau})^{\{k\}} \nonumber \\
&= -(f_{\tau})^i \hat{\Gamma}^j_{ij}
  -\sigma^i_{\{j\}\{k\}} (\partial_i \hat{\mathcal{R}}^{\{j\}})({f}_{\tau})^{\{k\}} +s_{\tau}.
\end{align}

Noting that $\sqrt{\hat{\gamma}} \hat{\mathcal{R}}^{\{r\}}
\hat{\mathcal{R}}^{\{\theta\}} \hat{\mathcal{R}}^{\{\varphi\}} =1$, the
evolution equation for the vector potential takes a particularly simple
form in our notation
\begin{equation}\label{eq:A_evol}
  \partial_t {A}_{\{i\}} = - {E}_{\{i\}}
  - \sigma^{l}_{\{i\} \{k\}} \hat{\mathcal{R}}^{\{k\}} \partial_l (\alpha \Phi - \beta^{\{j\}} A_{\{j\}}).
\end{equation}
The evolution equation of the EM scalar potential is given by
\begin{align}\label{eq:Aphi_evol}
\partial_t \hat \Phi &+ \sigma^i_{\{j\}\{k\}} \hat{\mathcal{R}}^{\{j\}} \partial_i ({f}_{\Phi})^{\{k\}} \nonumber \\
&= -\zeta \alpha \hat \Phi - (f_{\Phi})^i \hat{\Gamma}^j_{ij} -\sigma^i_{\{j\}\{k\}} (\partial_i \hat{\mathcal{R}}^{\{j\}})({f}_{\Phi})^{\{k\}},
\end{align}
and finally, the conserved rescaled magnetic field is calculated from
\begin{align}
  {\mathcal{B}}^{\{i\}} &= \sigma^{\{i\}\{m\}}_{l} \hat{\mathcal{R}}_{\{m\}}
\hat{\epsilon}^{ljk} \partial_j A_k \nonumber \\
  &= \sigma^{\{i\}\{q\}}_{n} \hat{\mathcal{R}}_{\{q\}} \hat{\epsilon}^{njk} \sigma^{\{l\}\{m\}}_k
  \Big({A}_{\{m\}} \partial_j \hat{\mathcal{R}}_{\{l\}}\nonumber \\
  &+ \hat{\mathcal{R}}_{\{l\}} \partial_j {A}_{\{m\}} \Big).
\end{align}

Introducing the following generalized sources
\begin{align}
  \Omega_{D} &\equiv -(f_{D})^i \hat{\Gamma}^j_{ij} -\sigma^i_{\{j\}\{k\}} (\partial_i \hat{\mathcal{R}}^{\{j\}})({f}_{D})^{\{k\}},\label{eq::source_dens}  \\
  (\Omega_{S})_{\{j\}} &\equiv \Big[- \sigma^{i}_{\{k\}\{l\}} \sigma^{\{m\}\{n\}}_q
  ({f}_{S})_{\{m\}}^{\{k\}} \left(\partial_i \hat{\mathcal{R}}^{\{l\}}_{\{n\}} \right) \nonumber \\
&- (f_{S})_q^i \hat{\Gamma}^k_{ik} + (f_{S})^i_k \hat{\Gamma}^k_{iq} + (s_{S})_q \Big] \sigma_{\{j\}\{p\}}^{q} \hat{\mathcal{R}}^{\{p\}} , \\
\Omega_{\tau} &\equiv -(f_{\tau})^i \hat{\Gamma}^j_{ij}
  -\sigma^i_{\{j\}\{k\}} (\partial_i \hat{\mathcal{R}}^{\{j\}})({f}_{\tau})^{\{k\}} +s_{\tau}, \\
  (\Omega_{A})_{\{i\}} &\equiv - {E}_{{\{i\}}} - \sigma^l_{\{i\}\{k\}} \hat{\mathcal{R}}^{\{k\}} \partial_l (\alpha \Phi - \beta^j A_j), \\
\Omega_{\Phi} &\equiv -\zeta \alpha \hat \Phi - (f_{\Phi})^i \hat{\Gamma}^j_{ij} -\sigma^i_{\{j\}\{k\}} (\partial_i \hat{\mathcal{R}}^{\{j\}})({f}_{\Phi})^{\{k\}}, \label{eq::source_Aphi}
\end{align}
we can write the reference-metric GRMHD evolution system in the following compact form suitable for
the integration in a finite volume scheme:
\begin{empheq}[box=\exactfbox]{align*}
  \partial_t D + \sigma^i_{\{j\}\{k\}} \hat{\mathcal{R}}^{\{j\}} \partial_i ({f}_{D})^{\{k\}} &= \Omega_{D},\\
  \partial_t {S}_{\{j\}}
  + \sigma^{i}_{\{k\}\{l\}} \delta^{\{n\}}_{\{j\}} \hat{\mathcal{R}}^{\{k\}} \partial_i \left(({f}_{S})_{\{n\}}^{\{l\}} \right)
&= (\Omega_{S})_{j},\\
  \partial_t {S}_{\{j\}}
  + \sigma^{i}_{\{k\}\{l\}} \hat{\mathcal{R}}^{\{k\}} \partial_i \left(({f}_{S})_{\{j\}}^{\{l\}} \right)
  &= (\Omega_{S})_{\{j\}},\\
  \partial_t \tau + \sigma^i_{\{j\}\{k\}} \hat{\mathcal{R}}^{\{j\}} \partial_i ({f}_{\tau})^{\{k\}}
&= \Omega_{\tau},\\
  \partial_t {A}_{\{i\}} &= (\Omega_{A})_{\{i\}}, \\
  \partial_t \hat \Phi + \sigma^i_{\{j\}\{k\}}
  \hat{\mathcal{R}}^{\{j\}} \partial_i ({f}_{\Phi})^{\{k\}}
&= \Omega_{\Phi}, \\
  {\mathcal{B}}^{\{i\}} = \sigma^{\{i\}\{m\}}_{l}
  \hat{\mathcal{R}}_{\{m\}}
  \hat{\epsilon}^{ljk} \partial_j A_k. &{}
\end{empheq}
We note that in Cartesian coordinates these equations reduce to the
equations of the {\it Valencia formulation} in a vector potential evolution.
In the following section, we will turn to the finite volume implementation of the
evolution system above, and in particular
how to deal with the rescaling factors multiplying divergences of fluxes in the
orthonormal basis with respect to the spherical background metric.

%
\subsection{Evolution equations in integral form}
\label{subsec:integral_form}
%
Systems of nonlinear hyperbolic partial differential equations (PDE)
such as the GRMHD evolution system presented above
are characterized by the fact that smooth initial
data can develop discontinuities in the variables
in finite time. The reason the evolution system is written in
conservative form is that, in such a form, a numerical scheme that
converges guarantees the correct Rankine-Hugoniot
conditions across discontinuities, which is called the
{\it shock-capturing} property. This property is
at the heart of high-resolution shock-capturing (HRSC)
methods that guarantee that the physics of the flow will
be correctly modeled by the numerical scheme in the
presence of discontinuities in the fluid variables.

Moreover, finite-difference schemes written in conservation
form guarantee that the convergence of the solution (if it exists)
will be to one of the {\it weak solutions} of the system of
PDEs~\cite{Lax1960}. Weak solutions are
characterized by being solutions to the integral form of the
conservation system.
The set of all weak solutions is too large to be of practical
use, as many (numerically) admissible weak solutions will
not represent physically relevant solutions.
Thus there is need for an additional (thermodynamic) condition,
the so-called {\it entropy condition} (namely that the entropy of
a fluid element must increase when crossing a discontinuity)
to guarantee that the numerical scheme will converge to the physical
solution.
The convergence of the numerical scheme is closely related
to its stability, and one useful measure is the {\it total-variation}
(TV) stability (see e.g.~\cite{Leveque1992} for a detailed
discussion).

Additionally, numerical schemes written in conservation form guarantee
that the conserved quantities of the system are numerically conserved
in the absence of sources or sinks. This means that the change of the state vector $U_A$
in time in a domain $\mathcal{V}$ that does not contain sources or sinks
will be given by the fluxes $F^i_A$ through the boundaries of the
domain $\partial \mathcal{V}$, a three-dimensional surface
which is defined as the standard-oriented hyper-parallelepiped
consisting of two spacelike surfaces $\{\Sigma_{x^0},\Sigma_{x^0+\Delta x^0}\}$
and the timelike surfaces $\{\Sigma_{x^i},\Sigma_{x^i+\Delta x^i}\}$
joining the two temporal slices together.

In a finite volume formulation, the evolution equations are integrated over the cell
volumes. For $D$, and similarly the other fluid variables, this amounts to the following
integrals that give the update of a conserved quantity in a given cell
(see, e.g.~\cite{Banyuls1997,FontLivingReview2008}):
\begin{equation}
  \partial_t (\langle D \rangle \Delta V) + \int_{\Delta V}
  \sigma^i_{\{j\}\{k\}} \hat{\mathcal{R}}^{\{j\}}
  \partial_i ({f}_{D})^{\{k\}} d^3x = \langle \Omega_{D}\rangle \Delta V.
\end{equation}
Here $d^3x = dr d\theta d\varphi$, and we have defined
\begin{eqnarray}
\langle D \rangle &=& \frac{1}{\Delta V} \int_{\Delta V} D d^3x, \\
\langle \Omega_{D}\rangle &=& \frac{1}{\Delta V} \int_{\Delta V} \Omega_{D} d^3x, \\
\Delta V &=&  \Delta r \Delta\theta \Delta\phi.
\end{eqnarray}
Notice the absence of the spherical background metric determinant $\sqrt{\hat{\gamma}}$
in the above expressions, as all knowledge about the spherical coordinates has been
moved to the background metric Christoffel symbols in the cell-centered source term
$\Omega_{D}$, together with our choice of $\bar{\gamma}=\hat{\gamma}$
\footnote{A different strategy could have been followed here, namely the integration
of Eq.~\eqref{eq:continuity_Dhat} in spherical coordinates directly using the
generalized Stokes theorem. This approach is followed in the {\tt CoCoNuT} code,
see~\cite{dimmelmeierPhD2001,CerdaCuranFluxCT2007} for details.}.
Up until now, the integration of the evolution equations over the cell volumes is exact.
Approximating the value of $D$ and the source $\Omega_{D}$ inside the cell volumes as
piecewise constant, and being equal to their cell-centered value (which is
a second order accurate approximation), we obtain
\begin{equation}
  \partial_t \langle D \rangle + \frac{1}{\Delta V} \int_V \sigma^i_{\{j\}\{k\}} \hat{\mathcal{R}}^{\{j\}}
  \partial_i ({f}_{D})^{\{k\}} d^3x = \langle \Omega_{D}\rangle.
\end{equation}
The integral $\int_V \sigma^i_{\{j\}\{k\}} \hat{\mathcal{R}}^{\{j\}}
\partial_i ({f}_{D})^{\{k\}} d^3x$ is not a true
divergence due to the appearance of the rescaling vector $\hat{\mathcal{R}}^{\{i\}}$,
and therefore we cannot use the divergence theorem to convert the volume integral into a
surface integral over the cell surface to arrive at finite volume scheme. We therefore
make a third-order approximation, setting $\hat{\mathcal{R}}^{\{i\}}$ to be piecewise constant and equal
to its value at the cell-center, denoted as $\langle \mathcal{R}^{\{l\}} \rangle_{ijk}$ (where the subscript $ijk$
denotes a cell), so that the volume integral can then be converted to a surface
integral of the fluxes through the cell faces,
\begin{eqnarray*}
  \sigma^l_{\{m\}\{n\}} \frac{\langle \mathcal{R}^{\{m\}}\rangle_{ijk}}{\Delta V} \int_V \partial_l ({f}_{D})^{\{n\}} d^3x\nonumber \\
  = \sigma^l_{\{m\}\{n\}} \frac{\langle \mathcal{R}^{\{m\}}\rangle_{ijk}}{\Delta V} \int_S {f}_{D}^{\{n\}} s_l dA,
\end{eqnarray*}
where $s_i$ is the outward pointing unit-normal to the cell surface, and $dA$
the surface element.
Therefore, all volume integrals in a finite volume scheme in the
reference-metric formalism will be \enquote{Cartesian}
in the sense that we integrate over \enquote{Cartesian} volumes and surfaces in
the spherical grid using fluxes in the orthonormal basis with respect to $\hat{\gamma}_{ij}$.
In our second-order accurate approximation, the surface integrals are given by
\begin{align}
  \frac{\langle \mathcal{R}^{\{r\}}\rangle_{ijk}}{\Delta V} \int_S {f}_{D}^{\{r\}} s_r dA &=
  \frac{\langle \mathcal{R}^{\{r\}}\rangle_{ijk}}{\Delta r} \Xi_r, \\
  \frac{\langle \mathcal{R}^{\{\theta\}}\rangle_{ijk}}{\Delta V} \int_S {f}_{D}^{\{\theta\}} s_{\theta} dA
  &= \frac{\langle \mathcal{R}^{\{\theta\}}\rangle_{ijk}}{\Delta \theta} \Xi_{\theta},  \\
  \frac{\langle \mathcal{R}^{\{\varphi\}}\rangle_{ijk}}{\Delta V} \int_S {f}_{D}^{\{\varphi\}} s_{\varphi} dA
  &= \frac{\langle \mathcal{R}^{\{\varphi\}}\rangle_{ijk}}{\Delta \varphi} \Xi_{\varphi},
\end{align}
where the symbols $\Xi_i$ are defined as
\begin{align}
  &\Xi_r \equiv \left(({f}_{D})^{\{r\}}_{i+\frac{1}{2},j,k}
  -({f}_{D})^{\{r\}}_{i-\frac{1}{2},j,k} \right), \\
  &\Xi_{\theta} \equiv \left(({f}_{D})^{\{\theta\}}_{i,j+\frac{1}{2},k}
  -({f}_{D})^{\{\theta\}}_{i,j-\frac{1}{2},k} \right ),\\
  &\Xi_{\varphi} \equiv \left(({f}_{D})^{\{\varphi\}}_{i,j,k+\frac{1}{2}}
  -({f}_{D})^{\{\varphi\}}_{i,j,k-\frac{1}{2}} \right).
\end{align}
We can then write the second-order accurate finite volume evolution equation for $D$ as
\begin{equation}
  \partial_t \langle D \rangle + \frac{\langle \mathcal{R}^{\{r\}}\rangle_{ijk}}{\Delta r} \Xi_r
  + \frac{\langle \mathcal{R}^{\{\theta\}}\rangle_{ijk}}{\Delta \theta} \Xi_{\theta}
  + \frac{\langle \mathcal{R}^{\{\varphi\}}\rangle_{ijk}}{\Delta \varphi} \Xi_{\varphi} = \langle \Omega_{D}\rangle,
\end{equation}
and similar for the Euler and energy equation.
The evolution equations for the vector potential~(\ref{eq:A_evol}) and
the EM scalar potential~(\ref{eq:Aphi_evol}) are treated differently:
the cell-centered electric field is calculated averaging the reconstructed
velocity and magnetic field at the surrounding cell faces (see Eq.~\eqref{Efield} below),
and the divergence in the EM scalar potential evolution equation is evaluated
using finite-differences.
This means that the time integration of the conserved
variables is given by the interface fluxes of matter and
energy-momentum of the fluid, as well as the (cell-centered)
sources. One can therefore approximate those {\it numerical} fluxes
(which depend on the solution at the cell interfaces) as the time-averaged
fluxes across cell interfaces during a time step.
In general, the approximation to the real solution on a
grid with finite resolution will be a piecewise continuous
function, which means that the fluxes can be obtained
by solving local Riemann problems at cell interfaces,
an idea first described by~\textcite{Godunov1959}.

Riemann problems are initial value problems (IVPs) with discontinuities
in the solution. During the evolution, a discontinuity in the fluid variables
decays into shock waves, rarefaction waves and
contact discontinuities. Shock waves move from the
higher to lower density regions, while rarefaction waves
move in the opposite direction. Contact discontinuities
are characterized by a discontinuity in the density,
while both pressure and velocity are constant
across them. In order to solve the Riemann problem,
we need to obtain the spectrum (eigenvalues and eigenvectors)
of the first-order system. The fluid data at the cell interfaces needed
to obtain the numerical fluxes via the solution of local Riemann
problems needs to be obtained from the cell averages.
A wide variety of higher order cell-reconstruction methods
are available in the literature (see e.g.~\cite{Toro2013}).
Regardless of their spatial order for smooth solutions, these reconstruction
techniques always reduce to first-order in the presence of
physical shocks and some reconstruction schemes even reduce to first order
at local extrema of the fluid variables
(such as the central density of a NS, for instance).

While the choice of variables is crucial to obtain the GRMHD
evolution equations in conservative form, it is usually the
primitive variables that are reconstructed at the cell interfaces.
To do this, one needs a conservative-to-primitive scheme, involving
numerical root finding.
Once we have obtained the numerical fluxes via the solution of
local Riemann problems, we update the solution of the
conserved variables by one time step with the numerical fluxes and
the sources. This is usually done employing high-order Runge-Kutta schemes~\cite{ShuOsher1988}.

From the structure of the equations in integral form, as noted above, we see that
they are \enquote{Cartesian} by virtue of having written the conservation laws in
the reference-metric formalism. The second-order accuracy is
achieved by encoding all the geometric information about the underlying coordinate
system in the cell-centered geometric source terms. Specifically, no care has to
be taken to distinguish the coordinate center and centroid of volume of the
computational cells. In general curvilinear coordinates, this is not the case,
as replacing the average value of a cell quantity with a point value is only
second-order accurate if the point is chosen to be the centroid of volume,
not the coordinate center~\cite{monchmeyer1989conservative,MignoneHO2014}.
To this end, the second-order accurate prescription
outlined could be applied to any existing Cartesian finite volume code by calculating the
appropriate reference-metric source terms and incorporating them in the time
integration of the evolution equations.

In the following section, we describe the necessary changes we performed
to enable the use of spherical coordinates via the GRMHD reference-metric
formalism presented above in \GRHydro~\cite{Baiotti2005,HawkeGRHydro2005,Loffler2012,GRHydro2014},
a publicly available GRMHD code that comes with the \ET.

%
\section{Implementation in the Einstein Toolkit}
\label{sec:SphericalGRHydro}
%
The \ET~\cite{ETK_2019_3522086} is an open source code suite for relativistic astrophysics
simulations. It uses the modular {\tt Cactus} framework~\cite{Cactus}
(consisting of general modules called \enquote{thorns})
and provides adaptive mesh refinement (AMR) via the {\tt Carpet}
driver~\cite{Carpet,Goodale2003,Schnetter2004}. In our vacuum implementation of
the BSSN equations~\cite{Mewes:2018szi} we have detailed how we enabled the
use of spherical coordinates in the \ET, having supplied our own
spacetime evolution thorn.

Enabling spherical coordinates in \GRHydro to arrive at a GRMHD code in spherical
coordinates in the \ET amounted to supplying a
different metric determinant and the appropriate reference-metric source terms,
so the changes to the existing code are minimally invasive and do not touch core
algorithms of \GRHydro.
Perhaps the most substantial change involved using the \NRPy
code~\cite{Ruchlin:2017com,web:NRPy} to replace the Cartesian GRMHD
source terms in \GRHydro with the generalized source terms
\eqref{eq::source_dens}--\eqref{eq::source_Aphi}. \NRPy converts these
expressions---written in human-readable, Einstein notation---into
optimized C-code kernels, automatically constructing finite-difference
derivatives at arbitrary order when needed. \NRPy was also used to fully
construct the C-code kernels for Einstein's equations written in both
BSSN (as described in~\cite{Mewes:2018szi}) and fCCZ4 formalisms.

At the interface between the spacetime and GRMHD evolution, the (physical) spacetime
variables $\alpha, \beta^i, \gamma_{ij}$, and $K_{ij}$ of the {\tt ADMBase} thorn
are passed to \GRHydro. As outlined above, rewriting the equations to evolve the
non-coordinate components of vectors, and dividing every $\sqrt{\gamma}$ by
$\sqrt{\hat{\gamma}}$ would amount to a great deal of rescaling and un-rescaling (both
vector components and determinants) in \GRHydro. Instead, we follow a different
route and pass the non-coordinate basis metric $\gamma_{\{i\}\{j\}}= e^{4 \phi}
(\delta_{\{i\}\{j\}}+h_{\{i\}\{j\}})$ and
shift $\beta^{\{i\}}$ as the {\tt ADMBase} variables in substeps of the method of lines
integration.
This means that $\sqrt{\det{\gamma}_{\{i\}\{j\}}}=e^{6 \phi}$ and raising and lowering indices of
non-coordinate vectors is achieved with the non-coordinate basis metric
$v_{\{i\}} = \gamma_{\{i\}\{j\}} v^{\{j\}}$.

As outlined above, given that the evolution equations in integral form
are \enquote{Cartesian}, the
different reconstruction methods that are available in \GRHydro
may be used {\it without modification}. These include:
total variation diminishing (TVD) with minmod; superbee~\cite{RoeSuperbee1986} and monotonized
central~\cite{vanLeer1977} limiters; the piecewise parabolic method (PPM)~\cite{ColellaPPM1984}
and its enhanced version that retains higher order at smooth extrema~\cite{mccorquodale2011,Reisswig2013};
monotonicity-preserving fifth order (MP5) reconstruction~\cite{SureshMP51997}; essentially non-oscillatory
reconstruction (ENO)~\cite{Harten1997}; as well as weighted
essentially non-oscillatory reconstruction
(WENO~\cite{shu1998essentially} and its variant
WENO-Z~\cite{Castrowenoz2011J}). Using these reconstruction
algorithms without modification would have been impossible had the
code been written in spherical coordinates without the reference-metric
formalism, as greater care must be taken when using these in spherical coordinates,
especially for higher order reconstruction methods (see e.g.~\cite{MignoneHO2014}).
We use the HLLE (Harten-Lax-van Leer-Einfeldt) approximate
Riemann solver~\cite{EinfeldtHLLE1988,HartenHLLE1983} present in {\tt
  GRHydro}, again without any changes to its implementation to
calculate the numerical fluxes through cell faces.

In order to achieve magnetic flux conservation, the cell-centered
electric field ${E}_{\{i\}}$ used in the update of the vector potential
is calculated as the average of the nonzero magnetic fluxes given by the
HLLE solver~\cite{TothDivB2000},
\begin{eqnarray}
  ({E}_{\{1\}})_{i,j,k} = &-& \frac{1}{4}\left(\alpha
  {v}^{\{2\}} {\mathcal{B}}^{\{3\}} - \alpha {v}^{\{3\}} {\mathcal{B}}^{\{2\}} \right)_{i,j-\frac{1}{2},k} \nonumber \\
  &-& \frac{1}{4} \left (\alpha
  {v}^{\{2\}} {\mathcal{B}}^{\{3\}} - \alpha {v}^{\{3\}} {\mathcal{B}}^{\{2\}} \right)_{i,j+\frac{1}{2},k} \nonumber \\
  &+& \frac{1}{4} \left(\alpha
  {v}^{\{3\}} {\mathcal{B}}^{\{2\}} - \alpha {v}^{\{2\}} {\mathcal{B}}^{\{3\}} \right)_{i,j,k-\frac{1}{2}} \nonumber \\
  &+& \frac{1}{4}\left(\alpha
  {v}^{\{3\}} {\mathcal{B}}^{\{2\}} - \alpha {v}^{\{2\}} {\mathcal{B}}^{\{3\}} \right)_{i,j,k+\frac{1}{2}}, \nonumber
\label{Efield}
\end{eqnarray}
and similarly for ${E}_{\{2\}}$ and ${E}_{\{3\}}$.
As shown in \cite{Giacomazzo:2010bx,Mocz2017}, the cell-centered
vector potential method we employ is identical to evolving the induction
equation directly with the so-called flux-CD scheme~\cite{TothDivB2000},
as the magnetic field is cell-centered and the curl of the
gradient $\hat{\mathcal{D}}_i \left(\alpha \Phi -\beta^j A_j \right)$ in the RHS
of $A_i$ is zero.

To mitigate high-frequency oscillations in the cell-centered vector potential
evolution, we add Kreiss-Oliger dissipation~\cite{kreiss1973methods} to the RHSs of
both $A_i$ and $\hat \Phi$~\cite{Etienne:2010ui,Giacomazzo:2010bx}.

One of the most delicate parts of GRMHD codes is the recovery of the primitive variables,
which usually requires nonlinear inversion. \GRHydro uses the conservative to
primitive routines scheme of~\cite{NobleHARM2006}. Some of the most problematic regimes for the
inversion are in regions of very high Lorentz factors and in magnetically dominated plasmas,
i.e. where $b^2/(2 P) \gg 1$. In those regions, the inversion errors may become comparable
to the truncation error and result in larger errors causing the evolution to fail eventually.
The biggest problem is the violation of physical constraints such as the positivity of $\rho$ and
$P$ during the recovery, as in this case the hyperbolicity of the evolution equations
breaks down~\cite{WuPPL2018}. As is customary in GRMHD codes, we use a tenuous atmosphere,
given that the GRMHD evolution equations break
down in true vacuum. The atmosphere region is particularly difficult to handle, as even
very small magnetic fields can result in very large ratios of magnetic to fluid pressure.
We use the following checks prior to primitive recovery:
\begin{enumerate}

\item In cells where $D<e^{6 \phi} \rho_{\mathrm{atm}}$, reset the cell to atmosphere assuming
a zero magnetic field (ignoring the contribution of the magnetic field to $\tau$)
and skip the primitive recovery. The magnetic field is fully evolved in the
atmosphere and always calculated from the curl of the vector potential.

\item Following~\cite{CerdaDuran:2008pv}, when a BH is present,
in regions where $b^2/(2 P)$ is greater than a user-specified threshold,
we raise the above criterion to reset to atmosphere, which avoids primitive inversion in
cells that are just above the atmosphere threshold. Effectively, this results in a higher-density
atmosphere, but in regions limited to high magnetic to fluid pressure ratios,
while allowing the use of a low-density atmosphere in regions of small magnetic fields.
This is important as a denser atmosphere can begin to affect the evolved physical system
of interest~\cite{ShibataAtmo2007}.

\item Following \cite{IGM2015} we check, after primitive recovery, whether $\rho$, $P$ and $W$ exceed
user-specified limits, and, if so, reset the primitives and then recalculate the conservatives.

\item Once an apparent horizon (AH) is found, we reset a small region
  deep inside the AH to atmosphere. In all the above steps, the
  magnetic field is never altered in any computational cell and always
  computed from the vector potential.

\end{enumerate}
In future versions of the code, problems related to the primitive recovery could be handled
by more modern algorithms, such as evolving the entropy $S$ and using it to recover the
pressure~\cite{NobleHarm2009}, or using different primitive recovery schemes,
see~\cite{SiegelC2P2018} for an overview. Another attractive approach could be
the use of physical-constraint-preserving methods~\cite{Wu2017,WuPPL2018,WuRMHD2018}.

A well-known problem of evolving hyperbolic PDEs in spherical coordinates is
the severe CFL limitation due to the nonconstant cell volumes in space, which become
smaller (therefore leading to smaller time steps) as the origin and axis are approached.
There are several approaches to mitigate this problem
(for an introduction, see e.g.~\cite{Boyd2001}), from various {\it multipatch}
approaches which remove the polar axis~\cite{Ronchi1996,GomezETH1997,Bishop1997,
Kageyama2004,Zink2008,Fragile2009a,Reisswig2013,Wongwathanarat2010,Melson2015,ShiokawaPW2018,BowenPWwave2020,Avara2020b},
mesh coarsening in the azimuthal direction at high latitudes~\cite{Liska2018},
radially dependent mesh coarsening in both polar and azimuthal angles~\cite{SkinnerFornax2019},
mesh coarsening as a conservative filter
operation~\cite{Mueller2015,Zhang2019,CoarseningCoconut2019,CoarseningAenus2019},
focusing resolution of the polar angle at the equator~\cite{Korobkin2011,Noble2012},
or the use of filters~\cite{Shapiro1970,Gent1989,Jablonowski2004}.

In order to avoid excessively small time steps in full 3D simulations, we employ
a radial and latitude dependent azimuthal fast Fourier transform (FFT) filter (using the {\tt FFTW3}
library~\cite{FFTW05}) that will be described
in detail in a companion publication~\cite{Zlochower2019}. In short,
we expand all evolved fields in the azimuthal direction in a Fourier series
and retain $m$-modes such that the time step at the pole is limited by
$n_{\varphi}=6$ points. Higher-order $m$-modes in the expansion are exponentially
damped, which is sufficient to prevent instability by violating the CFL condition.

We use the boundary condition thorn described in~\cite{Mewes:2018szi}. In summary,
internal boundary ghost zones for the $r$ boundary at the origin, and the $\theta$ and
$\varphi$ boundaries are copied in from points in the physical domain, accounting
for appropriate parity factors, which we list for rescaled and unrescaled vector
and tensor components in table~\ref{table:parities} for completeness.
The ghost zone to physical point mappings are as
follows:
\begin{itemize}
\item $r$-boundary at the origin:
\begin{eqnarray}
r & \to & - r \nonumber \\
\theta & \to & \pi - \theta, \nonumber \\
\varphi & \to & \varphi + \pi, \nonumber
\end{eqnarray}
\item
$\theta$-boundary at $\theta_{\rm min} = 0$:
\begin{eqnarray}
r & \to & r \nonumber \\
\theta & \to & - \theta, \nonumber \\
\varphi & \to & \varphi + \pi,\nonumber
\end{eqnarray}
\item $\theta$-boundary at $\theta_{\rm max} = \pi$:\footnote{We note
  there is a typo in the $\theta$ mapping of the $\theta_{\rm max} =
  \pi$ boundary in~\cite{Mewes:2018szi}, the correct mapping is the
  one shown here.}
\begin{eqnarray}
r & \to & r \nonumber \\
\theta & \to & 2 \pi - \theta, \nonumber \\
\varphi & \to & \varphi + \pi.\nonumber
\end{eqnarray}
\end{itemize}
Finally, ghost zones for $\varphi$ are set by imposing periodicity. We note
that our boundary condition requires an even number of grid points in the
$\varphi$ direction in order to ensure that ghost zones lie at the exact
locations of points in the physical domain.

\begin{table}\caption{Parity factors for rescaled and coordinate
vector and tensor components at the origin and polar axis.
The parity factors for contravariant components are the
same as for the covariant components shown.}
\begin{tabularx}{\columnwidth}{ZZZZ}
\toprule
&Origin &Axis \\
\hline
\noalign{\vskip 1mm}
$V_{\{r \}}$ &-- &+\\
$V_{\{ \theta \}}$ &+ &--\\
$V_{\{ \varphi \}}$ &-- &--\\
\noalign{\vskip 1mm}
\hline
\noalign{\vskip 1mm}
$V_r$ &-- &+\\
$V_{\theta}$ &-- &--\\
$V_{\varphi}$ &+ &+\\
\noalign{\vskip 1mm}
\hline
\noalign{\vskip 1mm}
$T_{\{r\} \{r\}}$ &+ &+ \\
$T_{\{r\} \{ \theta\}}$ &-- &-- \\
$T_{\{r\} \{\varphi\}}$ &+ &-- \\
$T_{\{\theta\} \{\theta\}}$ &+ &+ \\
$T_{\{\theta\} \{\varphi\}}$ &-- &+ \\
$T_{\{\varphi\} \{\varphi\}}$ &+ &+ \\
\noalign{\vskip 1mm}
\hline
\noalign{\vskip 1mm}
$T_{rr}$ &+ &+ \\
$T_{r\theta}$ &+ &-- \\
$T_{r\varphi}$ &-- &+ \\
$T_{\theta\theta}$ &+ &+ \\
$T_{\theta\varphi}$ &-- &-- \\
$T_{\varphi\varphi}$ &+ &+ \\
\noalign{\vskip 1mm}
\botrule
\end{tabularx}
\label{table:parities}
\end{table}

%
\section{Code tests}
\label{sec:code_tests}
%

In the following we show results for a series of code tests, ranging from special
relativistic test problems in a fixed background Minkowski spacetime to fully dynamical spacetime
evolutions of magnetized stable uniformly rotating neutron stars and the collapse of a magnetized
uniformly rotating neutron star to a Kerr BH~\cite{Kerr1963}.

%
\subsection{Tests in Minkowski spacetime}
%

The first set of tests is performed in a fixed background Minkowski
spacetime ($\{h_{\{i\}\{j\}}=\bar{A}_{\{i\}\{j\}}=0, K=\Theta=0,
\beta^{\{i\}}=B^{\{i\}}=\tilde{\Lambda}^{\{i\}}=0,
\alpha = e^{4 \phi} = 1 \}$), allowing us to compare the performance of the
MHD evolution with standard Newtonian tests. This enables us to
validate our implementation of the GRMHD evolution equations
with all metric terms set to flat space. To demonstrate
the code is working correctly in this setting, we show two tests
below. First, we solve a strong shock reflection problem without
magnetic fields (evolving pure HD problems by simply setting the
vector potential to zero everywhere initially). The second, more demanding test is an
explosion test problem. As shown below, these tests are do not exploit
symmetries of the spherical coordinate system, demonstrating that
the framework can, e.g., handle the passage of strong shocks through
the origin and polar axis.

%
\subsubsection{Relativistic spherical shock reflection test}
\label{subsec:RSSR}
%
Our first test is the relativistic spherical shock reflection
problem~\cite{1997ApJ...479..151M,Romero1996,GENESIS1999,Mignone2005,Nada2008}.
The test consists of an initially cold ($\eps \approx 0$)
fluid of unit density ($\rho=1$) flowing in uniformly with a
velocity of $v^r=v_{in}=-0.9$ towards the origin, where the fluid is compressed
and heated up resulting in a shock that travels upstream through the
inflow region. For numerical reasons, the problem is initialized with a small pressure
of $p=2.29\times 10^{-5} (\Gamma-1)$, where we use $\Gamma=4/3$.
The analytic solution to this problem is given by~\cite{1997ApJ...479..151M}:
\begin{equation}
\rho(r) = \begin{cases}
		\left(1+ |v_{in}| t / r \right)^2  & r > v_s t \\
		\left(1+ |v_{in}| / v_s \right)^2 \sigma & r < v_s t
\end{cases}
\label{eq:rho_RSSR}
\end{equation}
where the compression factor $\sigma$ and the shock velocity $v_s$ are given by
\begin{equation}
\sigma = \frac{\Gamma + 1}{\Gamma - 1}+\frac{\Gamma}{\Gamma-1} (W_{in} -1)
\end{equation}
and
\begin{equation}
v_s = \frac{\Gamma-1}{W_{in} + 1} W_{in} |v_{in}|,
\end{equation}
and where $W_{in}$ is the Lorentz factor of the inflowing fluid at the outer
boundary ($\approx 2.29$ for $v_{in}=-0.9$).

\begin{figure}
        \centering
        \includegraphics[width=\columnwidth]{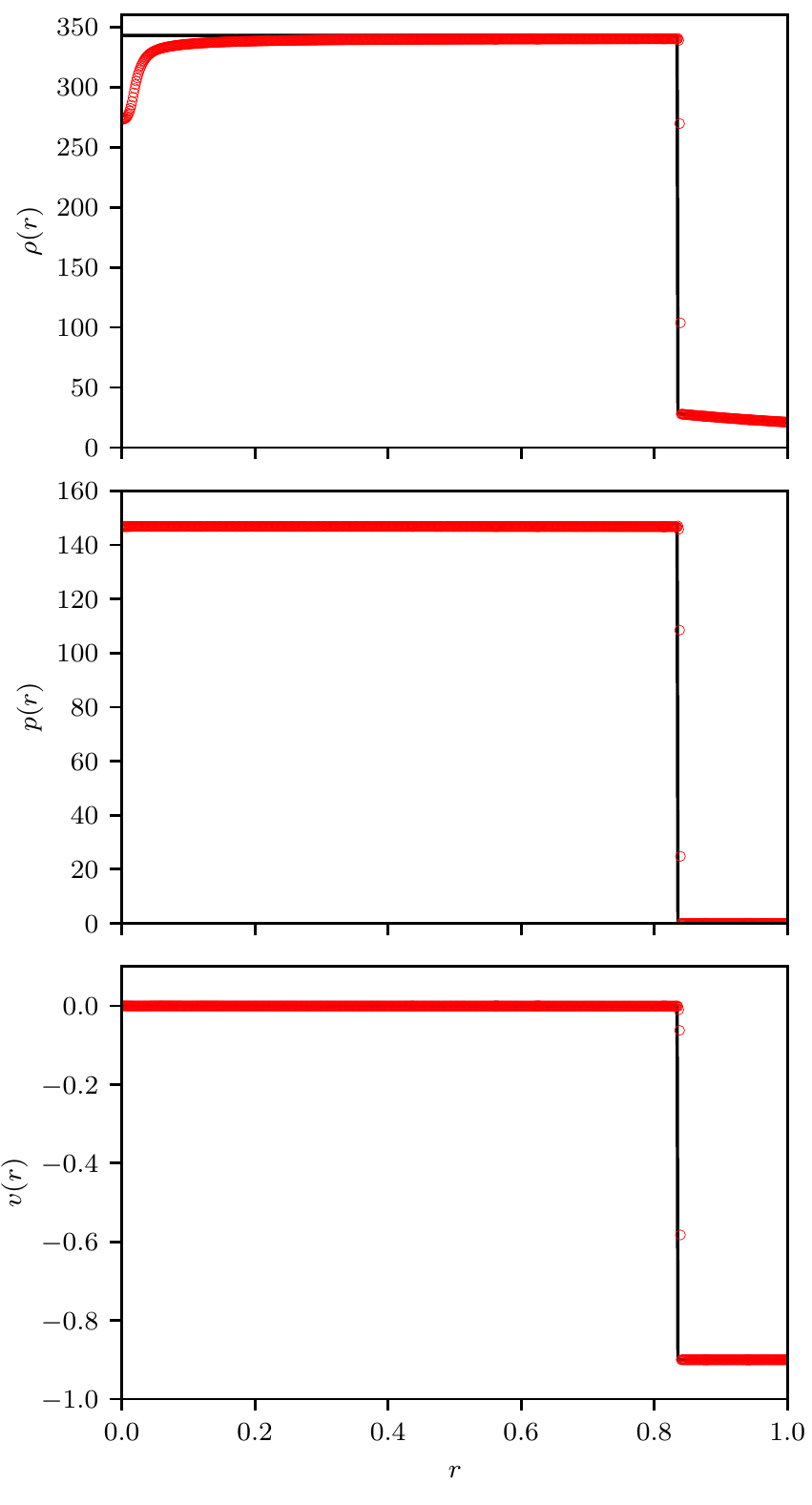}
        \caption{Radial density, pressure, and velocity profiles (from top to bottom)
        of the relativistic spherical shock reflection problem at time $t=4$. The red circles
        correspond to the numerical solution, while the analytic solution is shown as
        solid black lines.}
        \label{fig:RSSR}
\end{figure}

\begin{figure*}
\centering
\includegraphics[width=\textwidth]{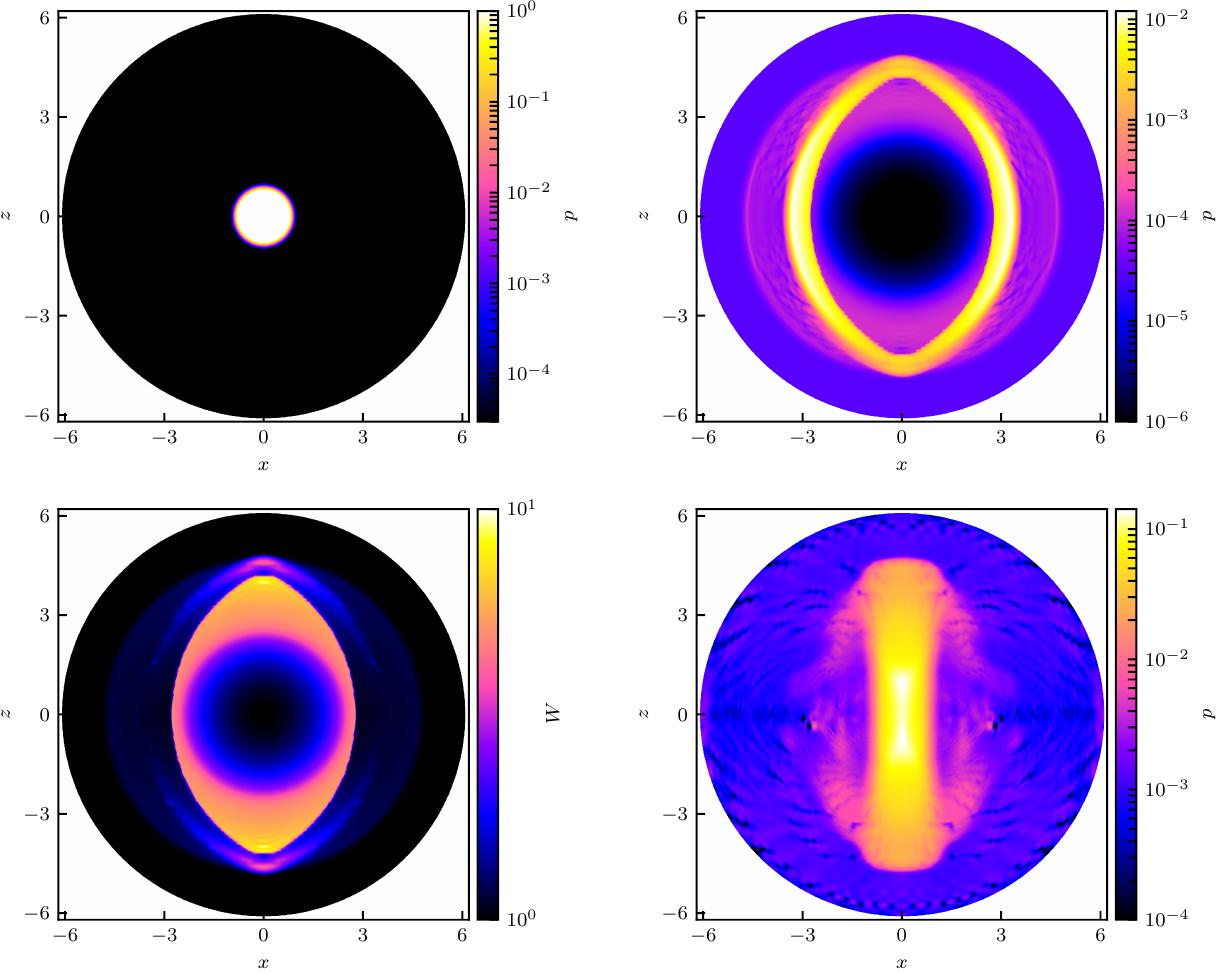}
\caption{Snapshots of magnetized spherical explosions in axisymmetry.
Top left: Initial pressure profile. Top right: Pressure at $t=4$, HLLE
Riemann solver and MP5 reconstruction. Bottom left: Lorentz factor at $t=4$, HLLE
Riemann solver and MP5 reconstruction. Bottom right:
Pressure at $t=4$, global Lax-Friedrichs fluxes and TVD reconstruction, initial $B^z=1.0$.}
\label{fig:spherical_explosion_axi}
\end{figure*}

Behind the shock wave $(r < v_s t)$, the fluid is at rest ($v^r=0$), and
internal energy is given by $\eps = W_{in}-1$.

For this test, we used 800 radial points in the interval [0:1], and
2 points in the $\theta$ and $\varphi$ directions, using the HLLE
Riemann solver, TVD reconstruction with a Minmod limiter and CFL
factor of 0.4. At the outer radial boundary, $\rho$ is set to the
analytic solution~\eqref{eq:rho_RSSR}, and $v^r$ and $p$ are kept
fixed at their initial values.
Figure~\ref{fig:RSSR} shows the radial profiles of
$\rho$, $P$ and $v^r$ of the numerical and analytic solution at $t=4$.

The global relative error at $t=4$ is $2.2\%$, $2.1\%$ and $1.2\%$, for $\rho$, $P$
and $v^r$, respectively. In the density profile, a significant drop near the origin is
present. This numerical effect is known as wall heating~\cite{Noh1987}, and seems to be
exacerbated in spherical coordinates due to the converging grid geometry~\cite{Rider2000}.

We also note that we observed significant postshock oscillations behind the
slowly moving shock when using higher order reconstruction methods. This appears to
be a known problem for HRSC schemes (see e.g.~\cite{stiriba2003numerical} and
references therein).

%
\subsubsection{Spherical explosion}
\label{subsec:spherical_explosion}
%

\begin{figure*}
\centering
\includegraphics[width=\textwidth]{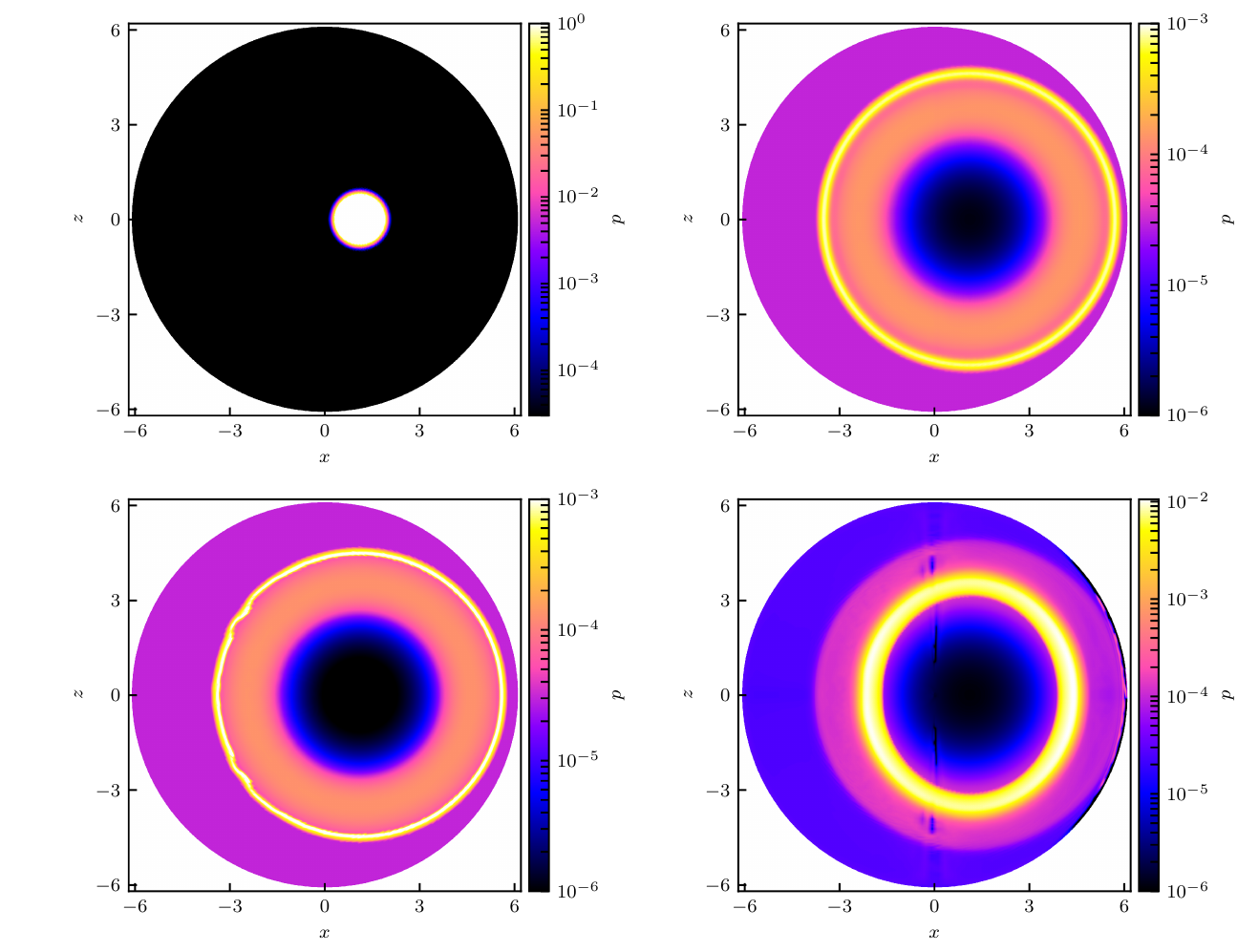}
\caption{Snapshots of off-centered spherical explosions.
Top left: Initial pressure profile. Top right: Pressure at $t=4$, HLLE
Riemann solver and TVD reconstruction, $B^i=0$. Bottom left: Pressure at $t=4$, HLLE
Riemann solver and MP5 reconstruction, $B^i=0$. Bottom right:
Pressure at $t=4$, global Lax-Friedrichs fluxes and TVD reconstruction with
initial magnetic field $B^z=0.1$ rotated by $45^{\circ}$ about the x-axis.}
\label{fig:spherical_explosion_o_t}
\end{figure*}

Next, we test the relativistic MHD evolution with a magnetized spherical explosion problem~\cite{CerdaDuran:2008pv}.
This test is the natural extension of the cylindrical explosion test proposed in~\cite{Komissarov1999}
to spherical coordinates, setting up spherically symmetric initial data.
Using the same jump conditions as~\cite{Komissarov1999}, the
initial data consists of an overdense ($\rho=\num{1e-2}$, $p=1.0$) ball of radius 1.0.
From a radius of 0.8 onwards, the solution is matched in an exponential decay to the
surrounding medium ($\rho=1 \times 10^{-4}$, $p=3 \times 10^{-5}$). The initial
pressure profile in the $y=0$ plane is shown in the top left panel of Fig.~\ref{fig:spherical_explosion_axi}.
We are using a $\Gamma-$law
equation of state (EOS) using $\Gamma=4/3$. In the magnetized case
the entire domain is initially threaded by a constant magnitude magnetic field parallel to the
z-axis ($B^z=0.1$), and the fluid velocity is set to zero everywhere in the domain initially.
As in the relativistic shock reflection problem, we use fixed background Minkowski spacetime
for this test problem.

We first model a magnetized spherical
explosion in axisymmetry, using
$(n_{r}=160,n_{\theta}=80,n_{\varphi}=4)$ points, with the outer
boundary $r_{\mathrm{max}}=6.0$. We use the HLLE Riemann solver and different reconstruction
methods for this test. The final distributions at $t=4$ for the pressure $P$
and Lorentz factor $W$ in the $y=0$ plane are shown in the top right
and bottom panels of Fig.~\ref{fig:spherical_explosion_axi}.

In the initially overdense explosion region, the fluid is only weakly magnetized, while being strongly magnetized
in the ambient medium. This results in a rich flow morphology in which the fast magnetosonic
wave travels out ahead in spherical symmetry at almost the speed of light, while the Alfv{\'e}n
wave shows a $\cos \theta$ dependence in propagation speed, traveling close to the speed
of light parallel to the initial magnetic field, while being significantly slowed down
in the direction perpendicular to the magnetic field, shown in the top right panel of Fig.~\ref{fig:spherical_explosion_axi} (see the discussion in~\cite{CerdaDuran:2008pv}).

During the explosion, the magnetic field is expelled from the initial explosion region, leaving a spherically symmetric
low-density region behind in which the fluid is at rest, as evidenced by the plot of the Lorentz factor
in the bottom left panel of Fig.~\ref{fig:spherical_explosion_axi}. The results seem to be
in very good qualitative agreement with the results presented in~\cite{CerdaDuran:2008pv}
(spherical coordinates and axisymmetry) and~\cite{Cipolletta:2019geh}
(Cartesian coordinates).

As a final axisymmetric test we perform the same explosion, but with an initial magnetic field
of much larger strength $B^z=1.0$. This is a very demanding
test for which we have used global Lax-Friedrichs fluxes and TVD reconstruction with the Minmod
limiter as they are more diffusive. In the bottom right panel of Fig.~\ref{fig:spherical_explosion_axi},
we plot the pressure distribution at $t=4$ in the $y=0$ plane for this test.
The morphology of the explosion changes completely and becomes bar-shaped,
as seen in Cartesian simulations of magnetized cylindrical and spherical
explosions (see e.g.~\cite{Beckwith2011,Cipolletta:2019geh}).
Compared to the more weakly magnetized explosion, more noise can be seen in the ambient region.
As discussed in ~\cite{Beckwith2011}, this test is most strenuous on the
conservative to primitive solver, so we believe that the noise is due to inversion failures.

Next, we test the code by modeling
an {\it off-center} spherical
explosion, both in relativistic HD and relativistic MHD. As this
test does not exploit the symmetries of our spherical coordinate
system, we perform it in full 3D. The initial data are identical to
the axisymmetric test described above, except the center of the explosion
region has been moved to $(x=1.1,y=0,z=0)$. The resulting initial
pressure profile in the $y=0$ plane is shown in the top left panel of Fig~\ref{fig:spherical_explosion_o_t}.
In addition, compared to the axisymmetric explosion, the initial magnetic field $B^z=0.1$
has been tilted by $45^{\circ}$ about the x-axis. This results in initial data that do not
reflect the symmetries of the spherical coordinate system at all.  For this full 3D test,
we use $(n_{r}=160,n_{\theta}=80,n_{\varphi}=160)$ points,
and use the azimuthal FFT filter to increase the time step to what it
would have been, had the simulation been performed with $n_{\varphi}=6$ points.

\begin{figure}
	\centering
	\includegraphics[width=\columnwidth]{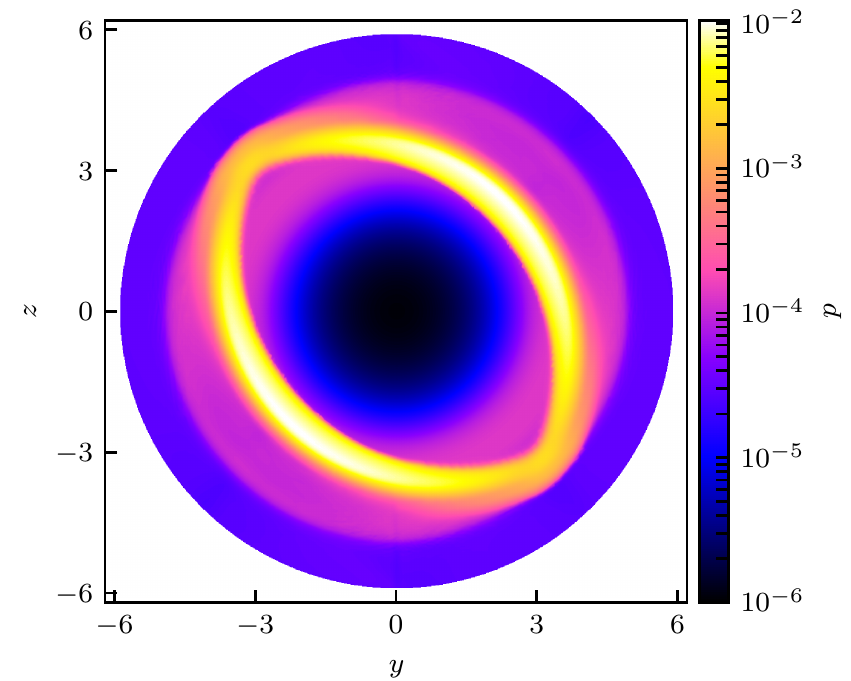}
	\caption{Pressure distribution in spherical off-centered explosion with a
	magnetic field initially tilted by $45^{\circ}$  about the x-axis.
	The plane shown is the y-z plane centered at the initial center of the explosion
	region $(x=1.1,y=0,z=0)$. Simulation performed with global Lax-Friedrichs fluxes and
	TVD reconstruction with Minmod limiter.}
	\label{fig:spherical_explosion_o_t_x11}
\end{figure}

We first perform two tests setting
the magnetic field to zero initially, using TVD and MP5 reconstructions. The results for the two different
 reconstruction schemes are
shown in the top right and bottom left panels of Fig.~\ref{fig:spherical_explosion_o_t},
respectively. When using TVD, there are no visible artifacts arising from the (in hydro only)
spherically symmetric shock passing through the origin and axis. The rarefaction region
is seen to be spherically symmetric as well, showing no artifacts. When using MP5, the shock width
is clearly reduced compared to TVD reconstruction, demonstrating a superior capture of the shock
with the higher order reconstruction, but small artifacts in those parts of the shock
that have passed the origin and polar axis can be seen. A similar test in hydro is presented
in~\cite{SkinnerFornax2019}. The bottom right panel of Fig~\ref{fig:spherical_explosion_o_t}
shows the final pressure distribution for the off-centered, tilted magnetic field spherical
explosion, which was performed using global Lax-Friedrichs fluxes and TVD reconstruction
with a Minmod limiter. This test displays more pronounced effects of the magnetized shock passing
through origin and polar axis, showing primitive recovery failures at the polar axis. The global
morphology is captured well nevertheless.

\begin{figure}
	\centering
	\includegraphics[width=\columnwidth]{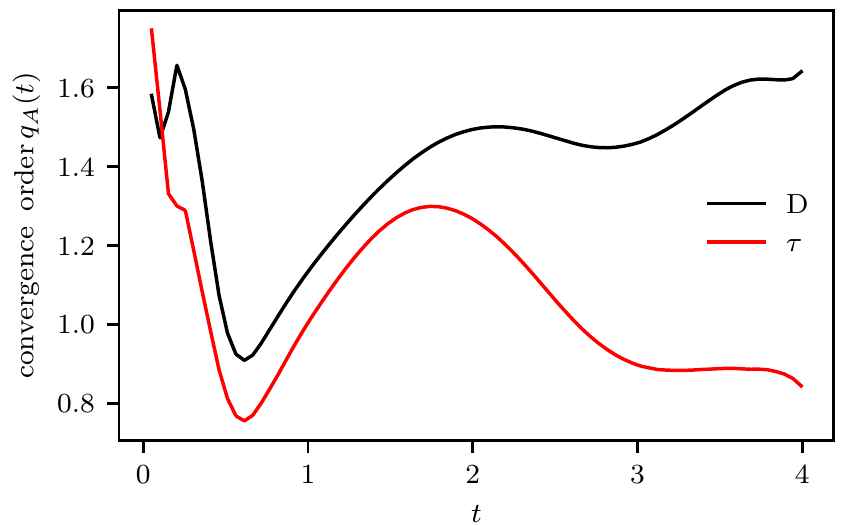}
	\caption{Convergence order in conservation of $D$ and $\tau$ for the
  spherical off-centered explosion with a magnetic field initially tilted
  by $45^{\circ}$  about the x-axis. See main text for details.}
	\label{fig:SE_o_t_convergence}
\end{figure}

It is worth noting that the numerical artifacts that can be seen in the
bottom panels of Fig~\ref{fig:spherical_explosion_o_t} are not concerning at all.
The test setup was deliberately chosen to push the code to its limits
by not exhibiting any (approximate) symmetries the framework was designed for,
and is a difficult test to pass even for Cartesian codes~\cite{Cipolletta:2019geh}.

Finally, in Fig.~\ref{fig:spherical_explosion_o_t_x11} we plot the final
pressure distribution in the $x=1.1$ plane, i.e. the plane vertically cutting through the initial
center of the explosion region. The flow morphology observed in the axisymmetric case
(top right panel of Fig.~\ref{fig:spherical_explosion_axi})
is clearly seen to be present and tilted by $45^{\circ}$, which is precisely the symmetry axis
picked out by the tilted magnetic field initially threading the computational domain.

As explained in Sec.~\ref{sec:summary} above, the conserved rest mass
density $D$ is not conserved to round off in our framework due to the appearance of geometric
source terms in the evolution equation for $D$~\eqref{eq:continuity_christoffel}.
Furthermore, the FFT filter we employ to circumvent the severe CFL limit in
full 3D simulations is inherently non-conservative as well. We therefore check
for the convergence of the total rest mass and total energy conservation ($s_{\tau}$,
the source term in the evolution equation for $\tau$~\eqref{eq:3d_energy_equation}
vanishes in Minkowski spacetime so that $\tau$ should be exactly conserved as well).
To do so, we calculate $\varepsilon_A(t)$, the volume integrated error arising
from nonconservation of a quantity $A=(D,\tau)$ at time $t$ as:
\begin{equation}
\varepsilon_{A}(t) = \int_{\Sigma} \left(A(t)-A(0)\right) \sqrt{\hat{\gamma}}\, dr d\theta d\varphi,
\end{equation}
and calculate the convergence order $q_A(t)$ as~\cite{BonaConvergence1998}:
\begin{equation}\label{eq:cf_known_result}
q_A(t) = \frac{1}{\ln(f)} \ln\left(\frac{||\varepsilon_{A}(t)||^{\mathrm{low}}}
{||\varepsilon_{A}(t)||^{\mathrm{high}}}\right),
\end{equation}
where $f$ is the ratio between the different resolutions used in the convergence test.
We calculate $q_A(t)$ in the most demanding variant of the spherical
explosion, the off-centered explosion with a tilted initial magnetic field, using
two different resolutions of $(n_r,n_{\theta},n_{\varphi})=(112,56,112)$ and (160,80,160),
corresponding to $f\approx \sqrt{2}$ and show the time evolution of $q_A(t)$
in Fig.~\ref{fig:SE_o_t_convergence}. In our numerical scheme, we would expect
the convergence order to be between 1 and 2, as our method is 2nd order accurate
while reducing to first order in the presence of shocks. The convergence order
of the conservation of total rest mass lies within that region, while for
the total energy it drops below first order at the time the numerical artifacts
at the polar axis seen in the bottom right panel of Fig.~\ref{fig:spherical_explosion_o_t}
start appearing. The maximum relative error in the conservation of total rest
mass and total energy in the high resolution test is $0.0016$ and $0.004$, respectively.

\begin{figure*}
\centering
\includegraphics[width=\textwidth]{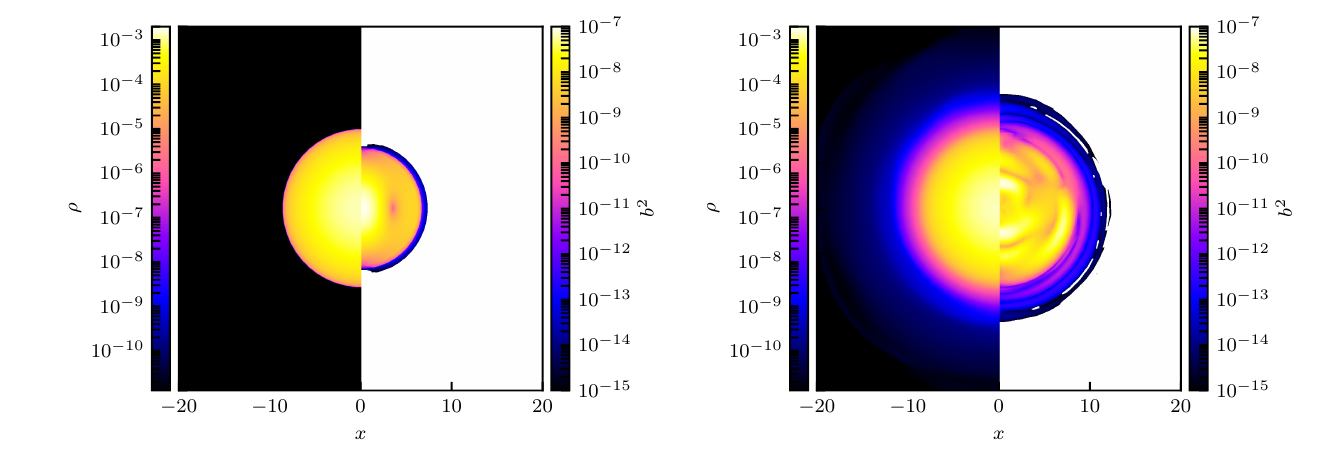}
\caption{$\rho$ and $b^2$ evolution of tilted model BU2, shown at $t=0$ (left) and
		 $t=25$ ms (right).}
		 \label{fig:BU2_xz_plane}
\end{figure*}

\subsection{Dynamical spacetime tests}

Next, we turn to dynamical spacetime evolutions of uniformly rotating neutron stars,
testing the coupled spacetime and GRMHD evolution of the {\tt SphericalNR}
framework. To test the framework in this regime, we evolve two uniformly rotating polytropes,
models BU2~\cite{StergioulasBU22004} and D1~\cite{Baiotti2005}, adding a weak
poloidal magnetic field initially. We perform tests of two important scenarios:
The long-term evolution of a stable equilibrium model, as well as the gravitational collapse of
a uniformly rotating polytrope to a BH. In the long-term evolution of model BU2, we initially tilt
the star's rotation axis by $90^\circ$ in order to test the evolution in full 3D without
symmetry assumptions. In this test, the fluid rotates through the polar axis during
the entire simulation, dragging the magnetic field with it through the polar axis
constantly. In the second test, we perform a simulation of the gravitational
collapse of model D1, testing all aspects of our framework: the correct coupled evolution
of the fCCZ4 and GRMHD equations leading to dynamical BH formation and ringdown to
a Kerr BH.

\begin{table}
\caption{Main properties of the relativistic polytrope models BU2~\cite{StergioulasBU22004}
and D1~\cite{Baiotti2005}:
central rest mass density $\rho_{c}$, rest- and gravitational masses $M_0$ and $M$,
the dimensionless angular momentum $J/M^2$, the circumferential stellar radius $R$,
the ratio of polar and equatorial radii of the star $r_p/r_e$, the ratio of
kinetic energy and gravitational binding energy $T/|W|$, the adiabatic
index $\Gamma$, the polytropic constant $K$, and the constants prescribing the
initial magnetic field $A_b$, $n_s$ and $P_{\mathrm{cut}}$ (see main text for details).}
\begin{tabularx}{\columnwidth}{ZZZ}
\toprule
\noalign{\vskip 1mm}
& BU2 &D1 \\
\hline
\noalign{\vskip 1mm}
$\rho_{c}$ 			&\num{1.28e-3} &\num{3.28e-3} \\
$M_0$				&1.58 &1.83 \\
$M$ 				&1.47 &1.67 \\
$J/M^2$				&\num{3.19e-1} &\num{2.07e-1} \\
$R$					&10.11 & 7.74\\
$r_p/r_e$			&0.9 & 0.95\\
$T/|W|$				&\num{2.44e-2} &\num{1.17e-2} \\
$\Gamma$ 			&2  & 2\\
$K$					&100  & 100\\
$A_b$				&2 & 1\\
$n_s$				&0 & 0\\
$P_{\mathrm{cut}}$	&\num{6.55e-6} & \num{4.25e-6}\\
\noalign{\vskip 1mm}
\botrule
\end{tabularx}
\label{table:NS_models}
\end{table}

\subsubsection{Tilted, magnetized, uniformly rotating neutron star}
\label{subsec:BU2}

The initial data are generated with the
{\tt RNS} code~\cite{StergioulasRNS1995}, which has been incorporated
as a thorn named \RNS in the \ET. As with all original Cartesian thorns
present in the \ET, in order to interface with {\tt SphericalNR}
we need to coordinate transform the Cartesian initial data generated
by the \RNS code and then rescale the evolved fields.
In order to test the FFT filter applied to both spacetime and GRMHD fields
and nontrivial dynamics in full 3D, we initially tilt the rotation axis of the
neutron star by $90^{\circ}$ about the $x$-axis so that the star's rotation axis is initially aligned
with the $y$-axis.
After generating the tilted fluid and spacetime data, we add a
small initial magnetic field, following the vector-potential-based
prescription of~\cite{Liu2008}:
\begin{eqnarray}
A_r &=& 0,\\
A_{\theta} &=& 0,\\
A_{\varphi} &=& A_{b} (r \sin \theta)^2 (1-\frac{\rho}{\rho_{c}})^{n_s} \mathrm{max}(P_{\mathrm{cut}}-P,0),\\
\Phi &=& 0,
\end{eqnarray}
where values of $A_b$, $\rho_c$, $n_s$ and $P_{\rm cut}$ are provided in Table \ref{table:NS_models}.
With this setup, the tilted, uniformly rotating star will constantly
drag the magnetic field through the polar axis during the evolution. While the initial data are polytropic,
we evolve the star with a $\Gamma$-law EOS.
We use a third order strong stability-preserving Runge-Kutta (SSPRK3)
method~\cite{ShuOsher1988,GottliebTVDRK1998}\footnote{Strong stability-preserving
time discretization methods have been called TVD methods historically~\cite{GottliebSSPRK2001}.}, implemented in the
{\tt MoL} thorn~\cite{Loffler2012}, the HLLE Riemann solver and WENO-Z reconstruction
for the simulations presented here. We have also tried ePPM and MP5 reconstruction, but
found a large symmetry breaking at late times when using those. To check for convergence of our code,
we evolve three different resolutions for 29 ms, which is more than 200 dynamical timescales of the star.

\begin{figure}
        \centering
        \includegraphics[width=\columnwidth]{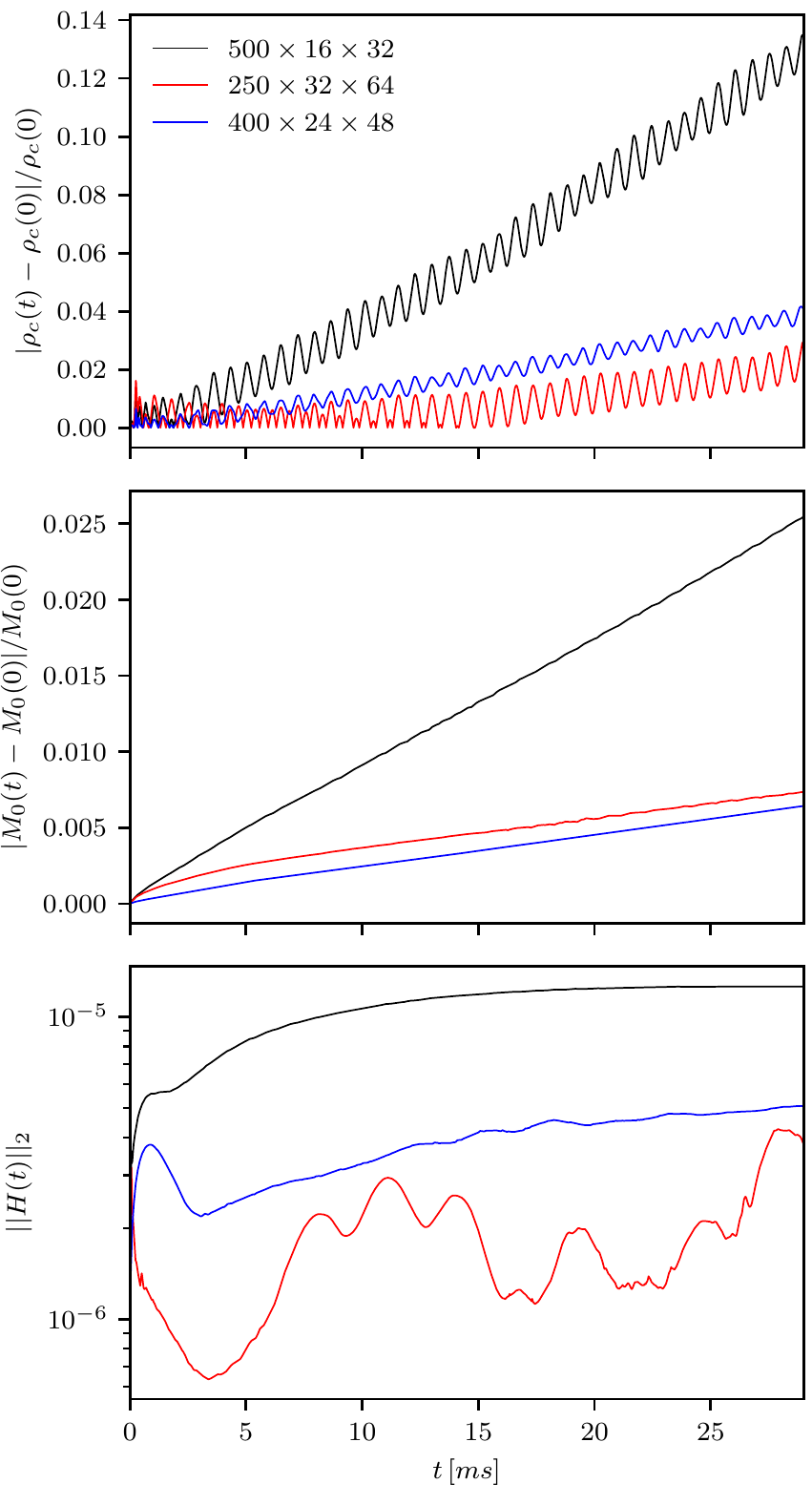}
        \caption{Evolution of central density $\rho_c$ (top panel), total rest mass
        $M_0$ (mid panel), and L2-norm of the Hamiltonian constraint (bottom panel)
        for magnetized model BU2. Three different resolutions are shown in each plot.}
        \label{fig:BU2_diagnostics}
\end{figure}

The $t=0$ and $t=25$ ms
distributions of $\rho$ and $b^2$ in the $y=0$ plane are shown in
Fig.~\ref{fig:BU2_xz_plane}. The star remains very stable and very contained, and there are no
large outflows from the stellar surface into the atmosphere, demonstrating the code's capability
to deal with the stellar surface. This is a difficult test, as the numerical
dissipation at the stellar surface is minimal in spherical coordinates, due to the
fact that the surface and computational cell surfaces are mostly
aligned (in Cartesian coordinates, this effect is seen along the coordinate axes,
see e.g. Fig. 3 in~\cite{RadiceHigherOrder2014}).
During the evolution, the quantity $b^2$ develops a richer morphology than it has in the
beginning, which we believe results from the fact that the initial poloidal field evolves into having
poloidal and toroidal components (while the initial data are uniformly rotating, the misalignment
between the star's rotation axis and the initial magnetic field dipole axis results in the generation of
a toroidal magnetic field.). In order to quantify the error arising from the FFT
filter and to check for the resolution dependence in the radial and angular
coordinates, we plot the following diagnostics in Fig.~\ref{fig:BU2_diagnostics}: the relative error
in central density and total rest mass in the top two panels, as well as the
evolution of the $L^2$-norm of the Hamiltonian constraint in the bottom panel.

\begin{figure}
        \centering
        \includegraphics[width=\columnwidth]{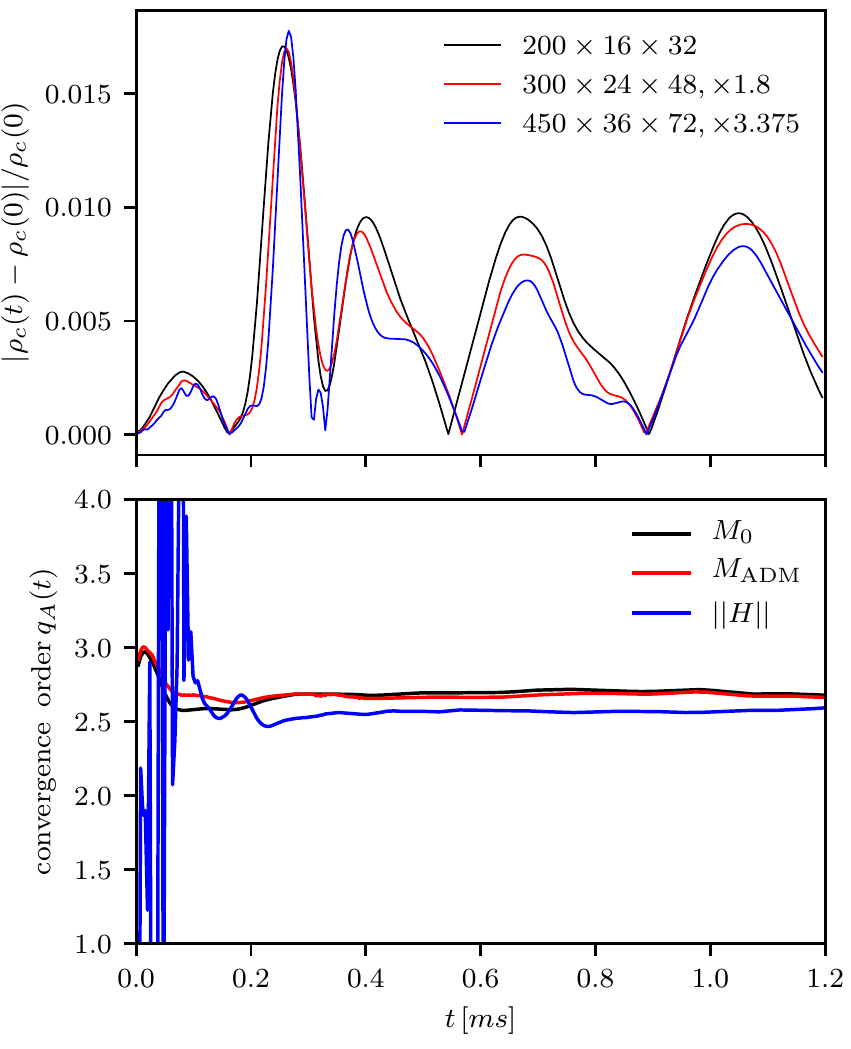}
        \caption{Convergence study for magnetized model BU2.
        Top panel: Relative error in $\rho_{c}(t)$ at multiple resolutions;
        medium and high resolution results are multiplied by the appropriate
        factor assuming a convergence order of 1.5. Bottom panel: Convergence order
        for the total rest mass $M_0$, ADM mass $M_{\mathrm{ADM}}$, and
        L1-norm of the Hamiltonian constraint.}
        \label{fig:BU2_convergence}
\end{figure}

\begin{figure}
        \centering
        \includegraphics[width=\columnwidth]{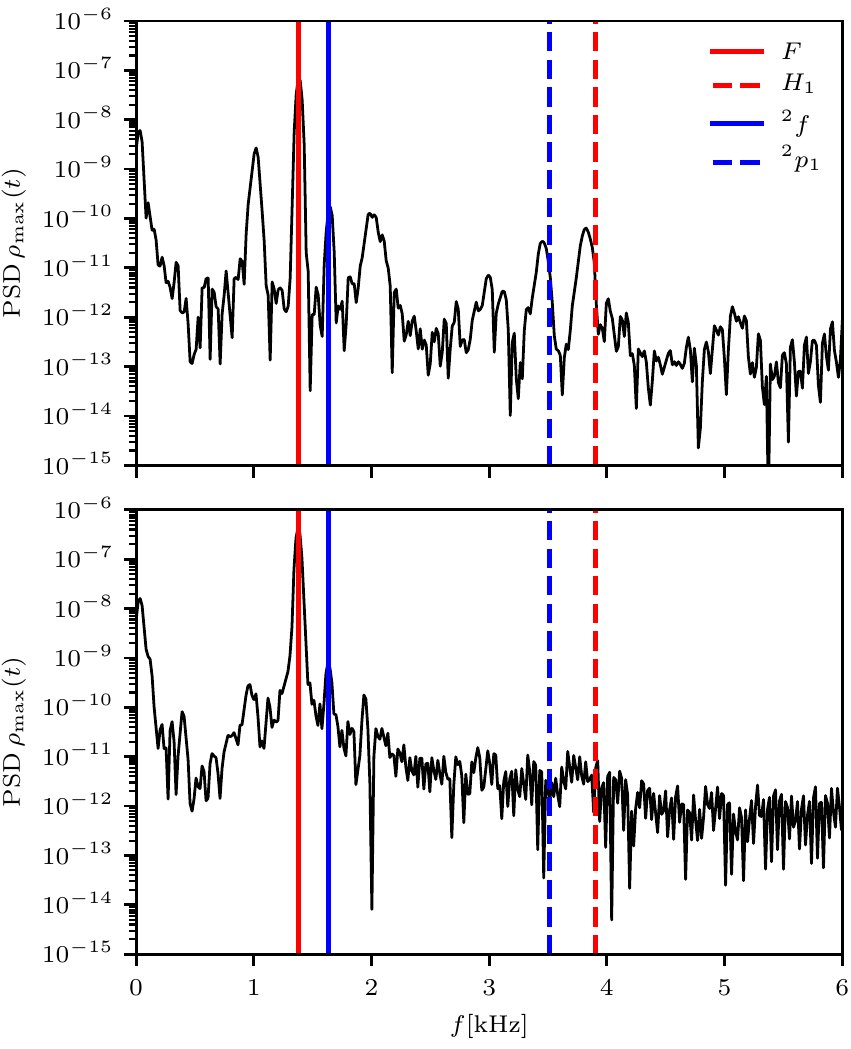}
        \caption{Power spectral density of $\rho_{\mathrm{max}}(t)$ evolution
        for resolutions of $(400 \times 24 \times 48)$ (top) and $(250 \times 32 \times 64)$
        (bottom) panel. The vertical lines indicate the
        fundamental $(F)$ and first overtone $(H_1)$ of the fundamental quasiradial
        ($l=0$) mode and the fundamental $(^{2}f)$ and first overtone $(^{2}p_1)$
        of the fundamental quadrupolar $(l=2)$ mode (see Table 5 of~\cite{DimmelmeierBU22006}).}
        \label{fig:BU2_rho_max_PSD}
\end{figure}

Additionally, in Fig.~\ref{fig:BU2_convergence} we show a convergence study
with three different resolutions, now increasing the resolution twice by a factor
of f = 1.5 in all three coordinate directions. The top panel shows the evolution of the error
in the central density evolution, where the medium and higher resolution errors
have been multiplied by 1.8 and 3.375, respectively, assuming a convergence order of 1.5.
Our code is formally second order, while reducing to first order in the presence of shocks.
The surface of the neutron star is a discontinuity, so we would expect the order
of convergence to be between 1 and 2.
The bottom panel shows the convergence factor for the total rest mass $M_0$~\eqref{eq:total_rest_mass},
the ADM mass evaluated as a volume integral, and the L1-norm of the Hamiltonian.
As in Sec.~\ref{subsec:spherical_explosion}, we calculate the convergence order
for $M_0$ and $M_{\mathrm{ADM}}$ using~\eqref{eq:cf_known_result}, while
the convergence order of the L1-norm of the Hamiltonian is calculated as~\cite{BonaConvergence1998}:
\begin{equation}
q_A(t) = \frac{1}{\ln(f)} \ln\left(\frac{||H(t)||^{\mathrm{low}}-||H(t)||^{\mathrm{med}}}
{||H(t)||^{\mathrm{med}}-||H(t)||^{\mathrm{high}}}\right).
\end{equation}

As the initial magnetic field is small, and the tilted rotation axis of the star
should not affect its dynamics during evolution, we calculate the frequencies of
oscillations in $\rho_{\mathrm{max}}(t)$ as a power spectral density for the two higher
resolution runs in Fig.~\ref{fig:BU2_rho_max_PSD}. We overlay the expected frequencies of
the fundamental quasiradial $(F)$ and quadrupolar $^{2}f$ modes and their
first overtones $(H_1$ and $^{2}p_1)$ (taken from Table 5 of~\cite{DimmelmeierBU22006}).
In both resolutions, the fundamental modes $F$ and $^{2}f$ are in very good agreement
with~\cite{DimmelmeierBU22006}, however the first overtones $(H_1$ and $^{2}p_1)$ of both
fundamental modes is not visible in the simulation with lower radial but higher angular
resolution, while being slightly shifted in the higher radial resolution simulation.

These tests show that our code is capable of evolving equilibrium neutron
stars with magnetic fields for many timescales, in a setup (constant fluid motion and
magnetic field dragging through the polar axis) that was chosen to be particularly
challenging for our framework.

%
\subsubsection{Collapse of a magnetized uniformly rotating neutron star}
\label{subsec:D1}
%

As our last test, we present a very important test problem for numerical
relativity simulations with matter: the collapse of a neutron star to a black
hole (see, e.g.~\cite{ShibataCollapse2000,Baiotti2005,BaiottiPRL2005,BaiottiCollapse2007,
Reisswig2013,DietrichCollapse2015}).
Using the \RNS code thorn in the \ET again, we setup the uniformly rotating polytrope
model D1~\cite{Baiotti2005} with a weak poloidal magnetic field added initially
and evolve its collapse to a Kerr BH. The initial data
specifications of this model are listed in Table~\ref{table:NS_models}.
The simulation is performed in axisymmetry, and the collapse is induced by
lowering the polytropic constant $K$ everywhere in the star by $2\%$ initially.
The simulation is performed using $(n_r=10000,n_{\theta}=32,n_{\varphi}=2)$ points, with the outer boundary
placed at $r_{\mathrm{out}}=500$. We evolve the conformal factor $W=e^{-2 \phi}$~\cite{Marronetti:2007wz},
use the SSPRK3 method for time integration, and the fCCZ4 damping parameters are set
to $\kappa_1=0.06,\kappa_2=0,\kappa_3=1$. We use WENO-Z reconstruction, the HLLE Riemann solver,
and a $\Gamma-$law EOS with $\Gamma=2$.

The atmosphere value for $\rho$ is set to be $10^{-8}$
times the initial density maximum $(\rho_{\mathrm{min}}=$\num{3.28e-11}. This simulation requires the use of the
higher atmosphere threshold in highly magnetized regions (described in Sec.~\ref{sec:SphericalGRHydro}
above), as the collapsing fluid leaves a highly magnetized atmosphere region behind.
In these regions, we reset a cell to atmosphere if $e^{6 \phi}\rho W < 100 e^{6 \phi} \rho_{\mathrm{min}}$.

We use the {\tt AHFinderDirect} thorn~\cite{Thornburg:2003sf, Schnetter:2004mc} to find
the AH~\cite{Thornburg2007a} once it has formed during collapse, and the
{\tt QuasiLocalMeasures} thorn~\cite{Dreyer:2002mx,Schnetter:2006yt} to calculate the
angular momentum of the AH during the evolution. The {\tt SphericalNR} interface to these Cartesian
diagnostic thorns in the \ET is described in~\cite{Mewes:2018szi}. The BH spin is measured using
a surface integral on the AH~\cite{Dreyer:2002mx} or the flat space rotational Killing vector
method~\cite{Campanelli2007,Mewes2015}.

\begin{figure}
        \centering
        \includegraphics[width=\columnwidth]{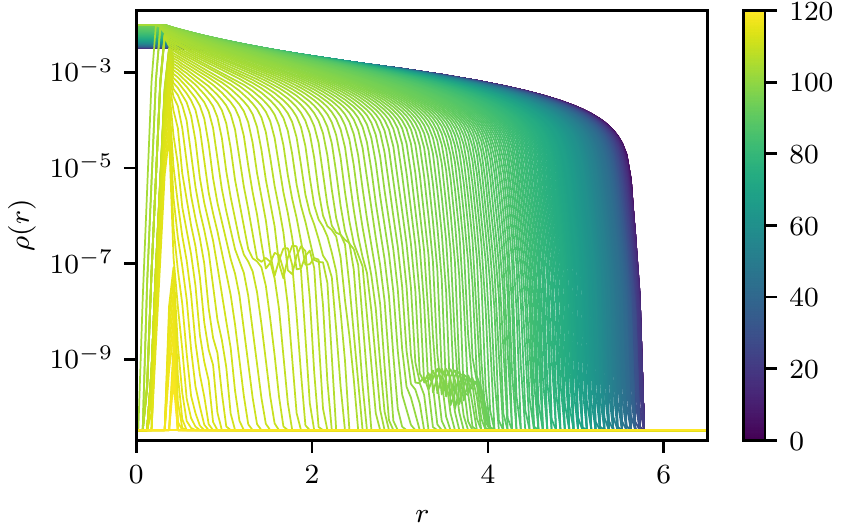}
        \caption{Temporal evolution of radial profiles of $\rho$ during
        the collapse of model D1, where the color bar indicates simulation time.}
        \label{fig:D1_collapse_rho}
\end{figure}

Deep inside the horizon, for points with coordinate radii $r < 0.2 \min(r_{\rm AH})$, we set
fluid variables to atmosphere values (the magnetic fields are evolved everywhere).
Figure~\ref{fig:D1_collapse_rho} shows the time evolution of radial profiles
(along $\theta=d\theta/2$) for the density  $\rho$. The evolution of $\rho$ shows our
modifications to the conservative to primitive solve in \GRHydro: As the star
collapses, $\rho$ is first capped to a maximum value, and then reset to atmosphere
deep inside the AH once it has been found (AHFinderDirect reports the first
finding of an AH at $t=103.8$).

\begin{figure}
        \centering
        \includegraphics[width=\columnwidth]{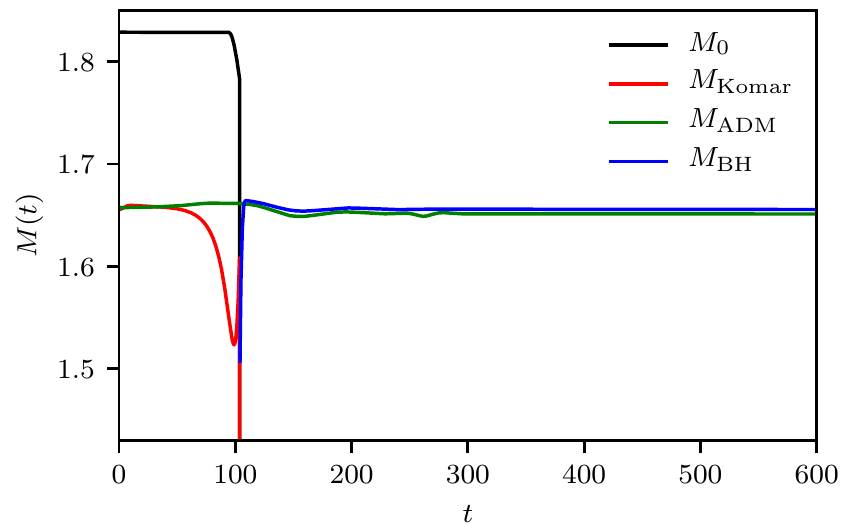}
        \caption{Time evolution of various mass measurements during the collapse
        of model D1.}
        \label{fig:D1_M}
\end{figure}

In Fig.~\ref{fig:D1_M} we plot different mass measurements for the duration of the
simulation.  Specifically, we monitor the total rest mass
\begin{equation}\label{eq:total_rest_mass}
M_0 = \int_{\Sigma_t} \rho W e^{6 \phi} \sqrt{\hat{\gamma}}\, d^3x,
\end{equation}
the Komar mass~\cite{Komar1959} evaluated as a volume integral
(see e.g.~\cite{Baumgarte2010})
\begin{equation}
M_{\mathrm{Komar}} = \int_{\Sigma_t} \left(\alpha (E + S)
- 2  \beta^i S_i  \right) e^{6 \phi} \sqrt{\hat{\gamma}} \, d^3x,
\end{equation}
and the ADM mass~\cite{Arnowitt2008} of the spacetime, evaluated as the sum from
contributions inside a finite radius $r_{\mathrm{in}}$, evaluated as a surface integral, and
those outside $r_{\rm in}$, evaluated as a volume integral~\cite{YoADMMass2002}
\begin{alignat}{2}\label{eq:adm_mass}
 M_{\mathrm{ADM}}
&= \frac{1}{16 \pi} \oint_{r_{\mathrm{in}}} &&(\Delta \Gamma^r - 8 e^{\phi} \bar{\mathcal{D}}^r \phi)
\, \sqrt{\hat{\gamma}} \, d\theta d\varphi \nonumber \\
&+ \frac{1}{16 \pi} \int_{r>r_{\mathrm{in}}} \Big[ &&e^{5 \phi} (16 \pi E + \bar{A}_{ij}\bar{A}^{ij} - \frac{2}{3}K^2) \nonumber \\
& \phantom{.} &&- \Delta\Gamma^{ijk}\Delta\Gamma_{jik} \nonumber \\
& \phantom{.} &&+ (1-e^{\phi})\bar{R} \Big] \, \sqrt{\hat{\gamma}} \, dr d\theta d\varphi.
\end{alignat}
Comparing with the expressions found in~\cite{YoADMMass2002} terms containing $\Delta \Gamma^k_{jk}$
are missing in the above expression, this is due to the fact that
\begin{equation}
\Delta \Gamma^k_{jk} = \frac{1}{\sqrt{\bar{\gamma}}} \hat{\mathcal{D}}_j \sqrt{\bar{\gamma}},
\end{equation}
which, together with our choice of $\bar{\gamma}=\hat{\gamma}$
results in $\Delta \Gamma^k_{jk}=0$.
Finally, we compute the BH mass calculated as the Christodoulou mass~\cite{Christodoulou1970}
\begin{equation}
M_{\mathrm{BH}} = \left(M^2_{\mathrm{irr}} + \frac{4 \pi J^2}{A} \right)^{\frac{1}{2}},
\end{equation}
where $M_{\mathrm{irr}}$ is the BH irreducible mass, $J$ the BH angular
momentum, and $A$ the AH area.

The mass measurements agree well with their initial value of the equilibrium
neutron star. As the collapse proceeds, the total rest mass $M_0$ is
seen to drop when the density deep inside the star is capped
(see Fig.~\ref{fig:D1_collapse_rho}) and then quickly drops to zero once an AH
has been found, as we exclude points within the horizon from volume integrals.
The same drop is observed in the calculation of $M_{\mathrm{Komar}}$, which also
exhibits a stronger deviation from its initial value earlier, due to the fact
that it is only defined for stationary spacetimes, and the collapse is an
inherently dynamical process. Towards the end of the simulations, the calculation
of the ADM mass via a surface integral shows oscillations related to the gravitational
radiation leaving the domain and being partially reflected at the outer boundary.

\begin{figure}
        \centering
        \includegraphics[width=\columnwidth]{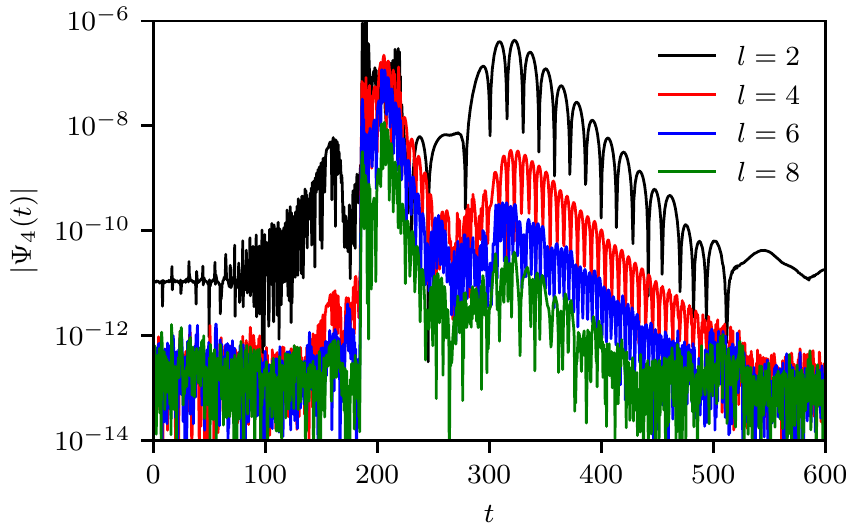}
        \caption{Even $l=2$ to 8, $m=0$ modes of the Weyl scalar $\Psi_4$.}
        \label{fig:D1_Psi4_even}
\end{figure}

\begin{figure}
        \centering
        \includegraphics[width=\columnwidth]{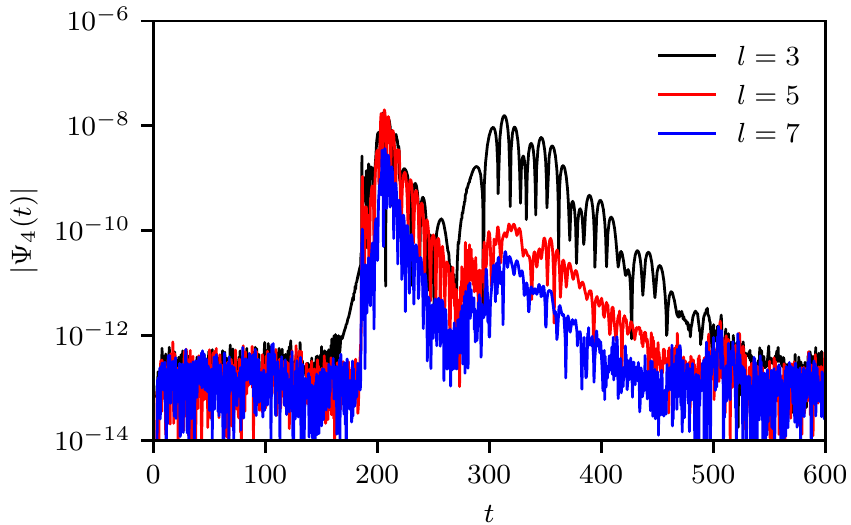}
        \caption{Odd $l=3$ to 7, $m=0$ modes of the Weyl scalar $\Psi_4$.}
        \label{fig:D1_Psi4_odd}
\end{figure}

After collapse, the newly formed BH is expected to quickly settle down to a Kerr BH
via the ringdown of the BH's quasinormal modes (for a review see~\cite{Berti2009}).
To see if our simulation reproduces this expected behavior, in Figs.~\ref{fig:D1_Psi4_even}
and~\ref{fig:D1_Psi4_odd} we plot the $l=2$ to 8, $m=0$ modes of the Weyl scalar $\Psi_4$,
split into even and odd $l$-modes, respectively. The ringdown of all modes is clearly seen,
as well beatings in the higher order modes, whose origin (the equal $m$ mode mixing of spherical
and spheroidal harmonics) we have explained in~\cite{Mewes:2018szi}.

The simulation shows that our spherical GRMHD code is capable of capturing the relevant
dynamics of the collapse of the magnetized uniformly rotating neutron star to a Kerr BH,
capturing the post-collapse ringdown to Kerr with very high accuracy, with all
modes dropping down to their initial background amplitudes.

%
\section{Conclusions and outlook}
\label{sec:conclusions}
%

We have extended our vacuum numerical relativity code in spherical coordinates
within the \ET~\cite{Mewes:2018szi} to a framework that
numerically solves the coupled fCCZ4/BSSN and GRMHD equations in spherical coordinates
without symmetry assumptions using a reference-metric formalism.

Extending the existing spacetime evolution thorn {\tt SphericalBSSN} to
evolve the fCCZ4 system with constraint damping as well, enables future
users of the framework with two distinct evolution systems for numerical relativity
in spherical coordinates. The spacetime evolution thorn was written
from scratch using \NRPy, while
the implementation of the reference-metric formalism
GRMHD equations derived in this work was built as an extension of the
\GRHydro thorn, again using \NRPy.

In our approach, the GRMHD equations in spherical coordinates acquire
a \enquote{Cartesian} form, as all information about the underlying
spherical coordinate system is encoded in source terms of the
equations.
This has allowed us to use many of the core Cartesian building blocks of the
HRSC finite volume implementation already present in \GRHydro without
modifications (and will enable the straightforward
inclusion of Cartesian finite volume building blocks such as more sophisticated
Riemann solvers in the future). Without the reference-metric approach, these
building blocks would need to be adapted to spherical coordinates.
Further, instead of evolving the magnetic field directly, the framework evolves
the cell-centered vector potential in the generalized Lorenz gauge, guaranteeing
the absence of magnetic monopoles to roundoff error during the evolution by
calculating the magnetic field as the curl of the vector potential.

We have tested our framework performing a set of demanding tests in flat background as well as
fully dynamical spacetimes. We have chosen setups where the symmetries of the
fluid are not aligned with the symmetries of the coordinate system
(counter to our original motivation for developing the code). These tests include
off-centered magnetized spherical explosions testing the passage of shocks and
rarefaction waves through the coordinate origin and polar axis, as well as
dynamical-spacetime simulations of a uniformly rotating neutron
star with its rotation axis misaligned with the polar axis of the
computational grid. Finally, we have shown that the code is able to
perform simulations of the collapse of a magnetized uniformly
rotating neutron star to a Kerr BH.

The {\tt SphericalNR} framework will be made public and proposed to be
included in a future official release of the \ET.

\begin{acknowledgments}
  The authors would like to thank the anonymous referee for useful comments
  and suggestions. We furthermore would like to thank Eirik Endeve for a
  careful reading of the paper, as well as Miguel {\'A}. Aloy, Mark J.~Avara, Dennis B.~Bowen,
  Pablo {Cerd{\'a}-Dur{\'a}n}, Isabel Cordero-Carri{\'o}n, Jos{\'e} A.~Font,
  Roland Haas, David Hilditch, Jos{\'e} M.~Ib{\'a}{\~n}ez, Kenta Kiuchi,
  Oleg Korobkin, Jens Mahlmann, Jonah M. Miller, Martin Obergaulinger,
  Scott C. Noble, David Radice, Ian Ruchlin, Erik Schnetter, and Masaru Shibata
  for useful discussions.
  We gratefully acknowledge the National Science Foundation (NSF) for
  financial support from Grant Nos.\  OAC-1550436, AST-1516150, PHY-1607520, PHY-1305730,
  PHY-1707946, and PHY-1726215 to Rochester Institute of Technology (RIT); PHY-1707526 to Bowdoin
  College; as well as OIA-1458952 and PHY-1806596 to West Virginia University. This work
  was also supported by NASA awards ISFM-80NSSC18K0538 and TCAN-80NSSC18K1488,
  as well as through sabbatical support from the Simons Foundation (Grant No.~561147 to TWB).
  V.M. was partially supported by the Exascale Computing Project (17-SC-20-SC),
  a collaborative effort of the U.S. Department of Energy (DOE) Office of Science
  and the National Nuclear Security Administration. Work at
  Oak Ridge National Laboratory is supported
  under contract DE-AC05-00OR22725 with the U.S. Department of Energy.
  V.M. also acknowledges partial support from the Spanish Ministry of Economy and Competitiveness (MINECO) through
  Grant No. AYA2015-66899-C2-1-P, and RIT for the FGWA SIRA initiative.
  This work used the Extreme Science and Engineering Discovery Environment (XSEDE)
  [allocation TG-PHY060027N], which is supported by NSF grant No. ACI-1548562, and the
  BlueSky and Green Prairies Clusters at RIT,
  which are supported by NSF grants AST-1028087, PHY-0722703, PHY-1229173, and PHY-1726215.
  Funding for  computer equipment to support the development of \SENR
  was provided in part by NSF EPSCoR Grant OIA-1458952 to West Virginia University.
  Computational resources were also provided by the Blue Waters sustained-petascale computing
  NSF project OAC-1516125. All figures in this paper were created using
  {\tt Matplotlib}~\cite{Matplotlib2007} for which we have used the {\tt scidata}~\cite{scidata:web}
  library to import {\tt Carpet} data.
\end{acknowledgments}

\bibliographystyle{apsrev4-1} \bibliography{references}

\end{document}